\documentclass[aps,prl,reprint,groupedaddress]{revtex4-2}

\usepackage{graphicx}
\usepackage{dcolumn}
\usepackage{bm}
\usepackage{braket}
\usepackage{color}
\usepackage{xcolor}
\usepackage{siunitx}
\usepackage{bibunits}
\usepackage{amsmath}

\begin{document}

\title{Intracycle Interference in Strong-Field Photoemission from a Tungsten Needle Tip}

\author{Timo Wirth\textsuperscript{1}, Jonas Heimerl\textsuperscript{1}, Stefan Meier\textsuperscript{1} and Peter Hommelhoff\textsuperscript{1,2}}
\affiliation{
 \textsuperscript{1}Department of Physics, Friedrich-Alexander-Universität Erlangen-Nürnberg (FAU), 91058 Erlangen, Germany\\
 \textsuperscript{2}Department of Physics, Ludwigs-Maximilians-Universität München (LMU), 80799 München, Germany}
 \email{Contact author: timo.wirth@fau.de}

\date{\today}

\begin{abstract}

Electrons photoemitted by strong laser fields can rescatter, generating attosecond high harmonics. Spectral peaks arise from inter- and intracycle interference of coherent electron wave packets. While intercycle interference is well known, intracycle interference of plateau electrons has not been observed. We present carrier-envelope phase–resolved photoemission spectra from a nanometric tungsten needle tip driven at 1570\,nm revealing such interference, supported by quantum orbit and TDSE simulations that elucidate attosecond-scale electron dynamics.

\end{abstract}

\maketitle

\begin{bibunit}[apsrev4-2]

When matter is exposed to intense light, it can enter the strong-field regime, leading to the generation of high harmonics and attosecond pulses of light~\cite{McPherson1987, Ferray1988, Brabec2000, LHuillier2024}. Similar strong-field phenomena occur at sharp metal needle tips, where an emitted electron can be driven back to the surface within the same optical cycle and elastically scatter, giving rise to the well-known rescattering plateau~\cite{Krueger2011, Herink2012, Krueger2012, Wachter2012, Krueger2018, Heimerl2023, Kim2023, Dienstbier2023_nature, Hergert2025}. The resulting strong-field photoemission spectra from metal tips closely resemble above-threshold ionization (ATI) spectra from atoms~\cite{Paulus1994, Becker2002}, as both originate from the same underlying mechanisms~\cite{Corkum1993}. The low-energy region of the spectrum arises from direct electrons that reach the detector without further interaction with the tip, whereas the plateau originates from rescattered electrons that are driven back to the surface and elastically scatter before detection.

The rescattering process in ATI and strong-field photoemission can be described classically within the three-step model, which can reproduce the formation of the plateau and its high-energy cutoff correctly~\cite{Corkum1993}. This model must be extended to track the electron phase to understand interference effects, i.e., effects resulting from the electronic matter wave nature. The most prominent feature of electron interference in photoemission spectra is the formation of peaks with an energy spacing of the photon energy of the driving laser. These peaks are often simply called photon order, multiphoton, or above-threshold ionization peaks. They result from interference of similar electron trajectories that reach the same final energies but are temporally spaced by one period (or an integer number of periods) of the light field and can be thought of as electron interference from slits in time~\cite{Lindner2005, Milosevic2006, Krueger2011}. Due to the temporal separation of the electron trajectories, this effect is also termed intercycle interference~\cite{Arbo2010}. Intercycle interference (photon) peaks have been observed in many photoemission spectra from tips~\cite{Schenk2010, Krueger2011, Bionta2013, Krueger2018, Heimerl2023} as well as in ATI spectra from atoms (see, e.g.,~\cite{Agostini1979, Paulus1994}).

\begin{figure}
\includegraphics[width=\columnwidth]{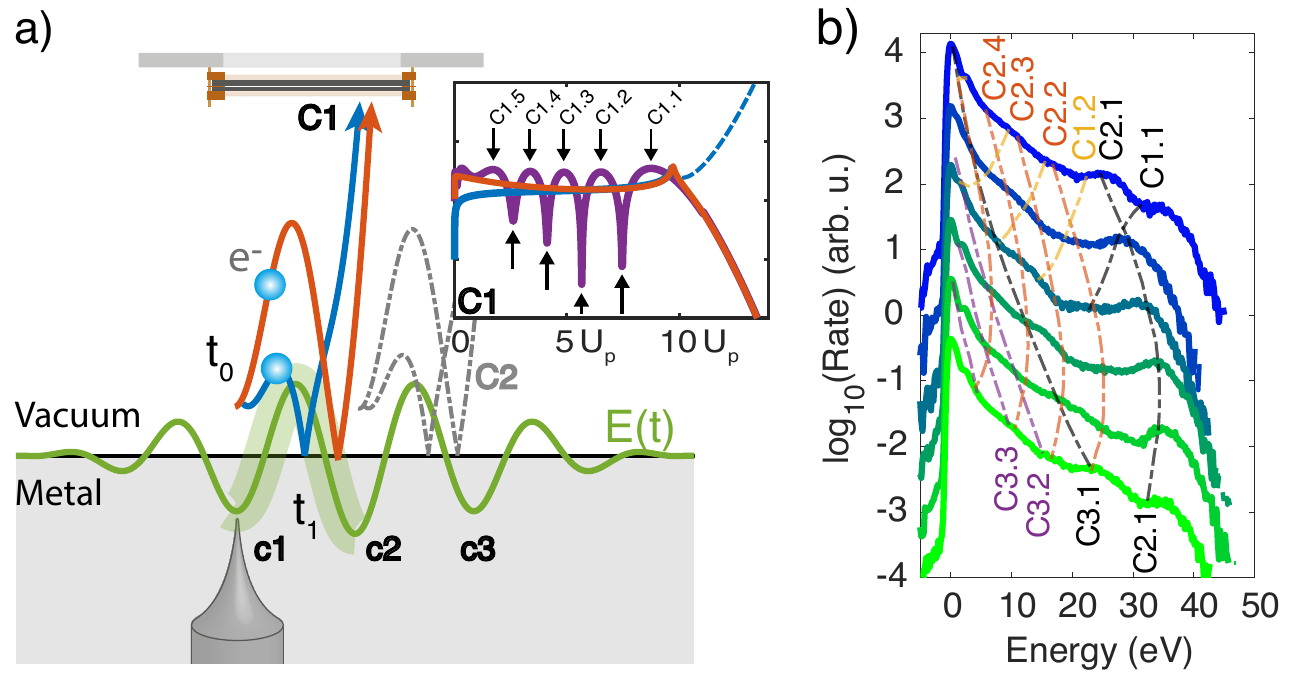}
\caption{\label{fig:Symbolic}Intracycle interference: conceptual picture and experimental spectra. (a) Schematic representation of intracycle interference in photoemission from a nanometric needle tip. Electrons born within the same laser cycle (c1, c2 or c3) and that have similar travel times may lead to intracycle interference. For ionization in each cycle there is a short (blue) and a long (red) orbit for every final energy. The isolated spectral contributions of both orbits have flat plateaus, as shown in the top right corner. Both orbits can interfere, leading to the modulation in their coherent sum (purple). The maxima and minima of this intracycle interference contribution are indicated with arrows. See Fig.~\ref{fig:Figure_2_QOVeranschaulichung} for more details. (b) Measurements of photoemission spectra for various CEPs (from top to bottom: 0, $0.375 \, \pi$, $0.750 \, \pi$, $1.125 \, \pi$, $1.5 \, \pi$, $1.875 \, \pi$). The dash-dotted lines highlight peaks resulting from intracycle interference. The energy resulting from the bias voltage of \SI{-30}{\volt} is subtracted and the spectra are shifted on the ordinate for clarity. The labels are introduced in the course of the text, see also Fig.~\ref{fig:Figure_2_QOVeranschaulichung}(b).}
\end{figure}

Generally, electron trajectories with temporal separations of less than one optical cycle that reach the same final energies can exist. The resulting interference is termed intracycle interference~\cite{Arbo2010}, illustrated in Fig.~\ref{fig:Symbolic}(a) for two rescattering electrons. Emitted at slightly different times with a temporal separation of $\sim$$300\,$as here (i.e., $\sim$6\,\% of the $5.2$\,fs long optical cycle), the electrons can reach the same final energies. As they travel along different paths to the detector, they accumulate different phases in the laser field, leading to varying electron matter wave interference at the detector. The periodic structures of the resulting intracycle interference patterns are approximately inversely proportional to the difference of the rescattering times and are hence substantially larger than for intercycle interference (photon orders)~\cite{Milosevic2006}. For this reason, intracycle interference is an ideal tool to probe electron coherence on much smaller timescales than possible with intercycle interference.

While intercycle interference in the form of photon order peaks has often been reported in photoemission from atoms and needle tips, intracycle interference has only been observed for direct electrons from atoms~\cite{Reichle2001, Reichle2003, Bergues2007, Kuebel2018} and has been found to be responsible for complex features in 2D photoemission spectra from dimers and molecules~\cite{Zhang2018, Khurelbaatar2024, Hasan2025}. These low-energy features result from all kinds of trajectories (see, e.g., Refs.~\cite{Gopal2009, Korneev2012, Hickstein2012, Möller2014, Maxwell2020} and references therein); in~\cite{Kuebel2021} a particular case of intercycle interference led to features in the rescattering plateau. Intracycle interference of rescattered electrons can be identified in a high number of \textit{simulated} strong-field spectra~\cite{Kopold2000, Kopold2001, Becker2002, Milosevic2006, Arbo2010} and is highly relevant for the precise shape of the plateau, however, it has, to the best of our knowledge, not been observed \textit{experimentally} in this energy region in any system thus far.

Here, we show strong-field spectra driven by two-cycle laser pulses emitted from a nanometric metal needle tip with plateaus up to 35\,eV and identify intracycle interference. By changing the carrier-envelope phase (CEP) of the pulses in the experiment, we observe a strong modulation of the plateau electrons. We employ two theory models, one based on solving the time-dependent Schrödinger equation (TDSE) and the other a quantum orbit simulation, which both show high agreement with the experiment. While TDSE simulations are well established for photoemission spectra from metal needle tips~\cite{Krueger2011, Dienstbier2023_nature}, quantum orbit theory has never been applied to this process. Mainly based on quantum orbit simulations, we can identify intracycle interference in the cutoff region here.

\section*{Experiment}

In the experiment, we use carrier-envelope phase-stable light pulses at a central wavelength of 1570\,nm with a pulse duration of 11.5\,fs (full width at half maximum, FWHM), equivalent to $\sim$$2.2$ optical cycles. We focus these pulses down to \SI{3}{\micro\meter} and align them on a tungsten needle tip with a radius of curvature of approximately \SI{10}{\nano\meter}, which is situated in an ultrahigh-vacuum chamber with a base pressure of $8 \cdot 10^{-10}$\,hPa. The optical electric field of the laser pulses is polarized parallelly to the symmetry axis of the tip. The electrons are emitted in a nonlinear photoemission process from the negatively biased tip apex ($U_{\mathrm{Tip}}=\SI{-30}{\volt}$) with an average rate of 0.4 electrons per pulse, which is low enough to rule out significant space charge effects on the overall form of the spectrum. We detect these electrons with a home-built time-of-flight spectrometer (for more details of the setup, see Ref.~\cite{Heimerl2025}).

Figure~\ref{fig:Symbolic}(b) shows energy spectra for various CEPs. The spectra clearly show a plateau and a strongly CEP-modulated cutoff position, as shown by the black dash-dotted lines (see explanation below). From the highest cutoff positions at 35\,eV and TDSE simulations (see below and Supplemental Material~\cite{SupplementalMaterial}), we can infer a near-field intensity of $I_{\mathrm{NF}} = 1.8 \cdot 10^{13} \, \mathrm{W/{cm}^2}$, including field enhancement at the tip. This value is slightly higher than expected from the 10-$U_{\mathrm{p}}$ law, since we take into account the quenching of the quiver motion in the optical near field of the metal needle tip~\cite{Herink2012, Heimerl2025}. This intensity yields a ponderomotive energy of 4.0\,eV and a Keldysh parameter of $\gamma \sim 0.9$. Further, we observe that the plateau region is not flat and shows humplike structures, clearly visible in Fig.~\ref{fig:Symbolic}(b). While the CEP-dependent cutoff modulations are well known from previous experiments and theory, we focus here on the precise shape of the spectrum close to the cutoffs, the humps in particular (see C2.1 and C3.1 in Fig.~\ref{fig:Symbolic}(b)).

\section*{Quantum orbit simulation}

For the interpretation of the experimental data, we apply quantum orbit theory for the rescattered electrons~\cite{Lewenstein1995, Kopold2000, Kopold2001, Becker2002, Milosevic2006}. Unlike a classical implementation of the three-step model, quantum orbit theory incorporates the quantum mechanical phase of electron trajectories and therefore correctly describes interference effects in photoemission spectra.

In quantum orbit theory, the spectrum is calculated from a set of contributions, the so-called saddle-point solutions. For the rescattered electrons, these are characterized by the triple $\left[t_0\left(\mathbf{p}\right), t_1\left(\mathbf{p}\right), \mathbf{k}\left(\mathbf{p}\right)\right]$, where $t_0\left(\mathbf{p}\right)$, and $t_1\left(\mathbf{p}\right)$ are respectively the instants of ionization and rescattering, $\mathbf{k}\left(\mathbf{p}\right)$ is the drift momentum between $t_0$ and $t_1$ and $\mathbf{p}$ the final momentum of the electron. The origin of each contribution, from which the spectrum builds up, can hence be straightforwardly explained with specific electron trajectories.

\begin{figure}[h!]
\includegraphics[width=\columnwidth]{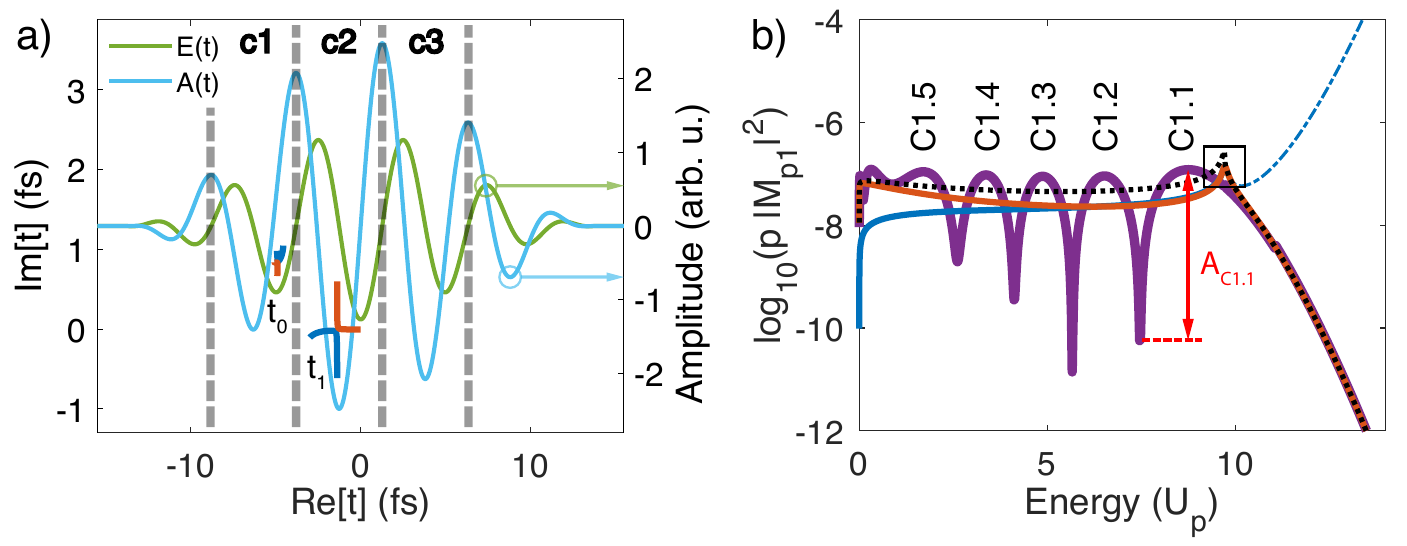}
\caption{\label{fig:Figure_2_QOVeranschaulichung}Quantum orbit theory and the origin of intracycle interference. (a): Two saddle-point solutions of the saddle-point pair C1 (ionization in laser cycle 1 (c1)) for various final energies, shown for $\mathrm{CEP} = 0$ and the parameters of the experiment. The associated optical electric field and vector potential are shown in green and blue. The saddle-point solutions constitute short (blue) and long (red) orbits. (b) Spectral yields of the short and long orbits and of their coherent sum. While the isolated spectral contributions $p {|M_{p1}|}^2$ of short and long orbits result in flat plateaus (dotted black line), their coherent sum (purple) shows a modulation with several humps. This modulation is due to interference of the two discussed trajectories, hence intracycle interference. There are similar contributions C2 and C3 with ionization times $t_0$ in laser cycle 2 (c2) and 3 (c3), from which the resulting full spectrum predominantly builds up. The intracycle interference maxima are labeled starting from the cutoff C1.1 toward lower energies. It should be noted that the sharp peaks at the $10$-$U_{\mathrm{p}}$~cutoff (square) of the isolated contributions (red and blue lines) and their incoherent addition (dotted black line) are artificial and not present in their coherent addition (purple line), for which the uniform approximation was used. An extended version of this figure can be found in Fig.~\ref{fig:Figure_2_QOVeranschaulichung_EM} in the End Matter.}
\end{figure}

In Fig.~\ref{fig:Figure_2_QOVeranschaulichung}(a), two saddle-point solutions that lead to intracycle interference are shown for a set of final energies $E=p^2/{\left( 2 m\right)}$, namely the red and blue curves, as functions of the real time (horizontal axis) and the imaginary time (vertical axis). The imaginary time components are mathematically necessary to fulfill the saddle-point equations, especially Eq.~(18) in Supplemental Material~\cite{SupplementalMaterial}. The real and imaginary times of the saddle points determine the quantum mechanical phase and spectral weight of the corresponding quantum orbit via the action term $\exp{\left[ i S\left( t_{0}, t_{1}, \mathbf{k}\right)/\hbar \right]}$, respectively. The two saddle-point solutions represent short (S, blue line) and long (L, red line) orbits and form a saddle-point pair.

In Fig.~\ref{fig:Figure_2_QOVeranschaulichung}(b), we show the spectral weights of both short (blue) and long (red) orbits as a function of the final energy. Both saddle-point solutions have an almost flat plateau up to an energy of approximately $10 \, U_\mathrm{p}$, similar to classical expectations. After the cutoff, the spectral weight of the short orbit (blue) exponentially increases, which is a commonly known artifact~\cite{Becker2002, Milosevic2006}, and which we make sure is always suppressed, either manually or by use of the uniform approximation (UA~\cite{Faria2002, Milosevic2007CCs, Milosevic2014}). When we sum the two complex solutions coherently (purple; $p \left(|M_{\mathrm{S}}+M_{\mathrm{L}}|^2\right)$ in terms of UA), a strong modulation is visible, approximately spaced by $1.6 \, U_{\mathrm{p}}$, while in the incoherent sum (black dotted line; $p \left(|M_{\mathrm{S}}|^2+|M_{\mathrm{L}}|^2\right)$), the plateau appears flat. Crucially, we find the appearance of a local maximum in the spectrum at $8.75 \, U_{\mathrm{p}}$. We emphasize that this quantum orbit spectrum results from just two trajectories, both starting and rescattering within the same laser cycles. The resulting interference phenomenon -- hence clearly represents intracycle interference. Most notably, this leads to a hump close to the 10-$U_{\mathrm{p}}$~cutoff. Without intracycle interference, the spectrum would be flat throughout. See end matter and supplement~\cite{SupplementalMaterial} for more details.

Whereas in Fig.~\ref{fig:Figure_2_QOVeranschaulichung} only two saddle-point solutions are shown, we calculate the full spectrum from the coherent sum $p |\sum_{i=1} M_i|^2$ of all saddle-point solutions that are physically relevant. Here, these are the quantum orbits with ionization times $t_0$ in the three main cycles of the pulse (denoted as c1, c2, and c3 in Fig.~\ref{fig:Figure_2_QOVeranschaulichung}(a)) and travel times $t_1-t_0$ of less than one optical period, i.e., the ones of Fig.~\ref{fig:Figure_2_QOVeranschaulichung} and their analogs in cycles 2 and 3 -- all others have negligible spectral weight. The quantum orbits are labeled C1, C2, and C3 according to the laser cycle (c1, c2, or c3) in which ionization occurs (see Fig.~\ref{fig:Figure_2_QOVeranschaulichung}(a)). Furthermore, also the maxima of the isolated saddle-point pair contributions (intracycle interference maxima) are labeled, starting from the highest energy maximum toward lower energies (see Fig.~\ref{fig:Figure_2_QOVeranschaulichung}(b)). In this notation, C[X].1 denotes the highest energy maximum of the contributions from cycle [X]. C[X].2 and C[X].3, respectively, point to the second and third maximum of the contribution from cycle [X] toward smaller energies (see Fig.~\ref{fig:Figure_2_QOVeranschaulichung}(b)). We note that the incoherent addition of the isolated pair contributions $p \left(|M_{\mathrm{C1S}}+M_{\mathrm{C1L}}|^2+|M_{\mathrm{C2S}}+M_{\mathrm{C2L}}|^2+\dots\right)$ matches the spectral structure of the coherent addition of all contributions $p \left(|M_{\mathrm{C1S}}+M_{\mathrm{C1L}}+M_{\mathrm{C2S}}+M_{\mathrm{C2L}}+\dots|^2\right)$ remarkably well, except for the presence of photon order peaks (see Fig.~\ref{fig:Figure_3_240125_1459}(b) below and Supplemental Material~\cite{SupplementalMaterial} for more details). To explain the structure of spectra, it is therefore sufficient to plot the isolated pair contributions C1, C2, and C3 (see Figs.~\ref{fig:Figure_3_240125_1459}(b) and~\ref{fig:Figure_3_240125_1459}(c)).

\begin{figure}[h!]
\includegraphics[width=\columnwidth]{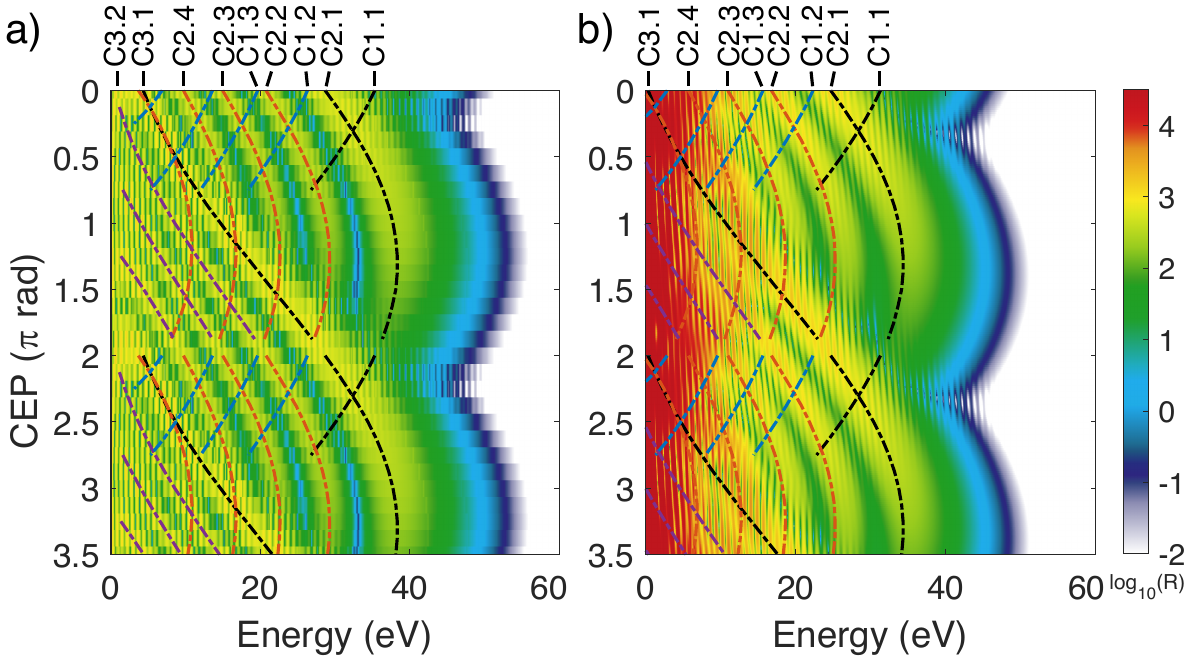}
\caption{\label{fig:Fig1_Colormaps} Intracycle interference in simulated photoemission spectra. CEP map resulting from a quantum orbit simulation (a) and from a TDSE simulation (b), both for a near-field intensity of $I_{\mathrm{NF}} = 1.8 \cdot 10^{13} \, \mathrm{W/{cm}^2}$ and a pulse duration of 10\,fs. The isolated intracycle interference maxima (see Fig.~\ref{fig:Figure_2_QOVeranschaulichung}(b)) from the quantum orbit simulation (curves in (b)) are plotted in (a) with an energy shift of 4.0\,eV to account for the optical near field of the tip (see Supplemental Material~\cite{SupplementalMaterial}). This allows for a detailed interpretation of the TDSE result and the measurement (see Fig.~\ref{fig:Symbolic}(b)). In both simulations, also peaks spaced finely by the photon energy ($\sim$$0.8\,$eV) are observed, resulting from intercycle interference. We note here already that the simulations match the measurement (see Figs.~\ref{fig:Figure_3_240125_1459} and~\ref{fig:Symbolic}(b)) perfectly. See text for details.}
\end{figure}

In Fig.~\ref{fig:Fig1_Colormaps}(a), we show a CEP map based on quantum orbit simulations for the rescattered electrons. We observe a CEP-dependent modulation of the yield spaced by approximately 6\,eV, highlighted by the dash-dotted lines. These modulations represent the above-discussed maxima C[X].[Y], forming arches in the CEP map. They originate from the intracycle interference maximum [Y] of the pair contribution from cycle [X]. We see that the plateau region is mostly dominated by the pair contributions of cycle 2 (C2) and cycle 3 (C3). We note that the cycle names always refer to the CEP in the range of $[0; 2 \, \pi[$.

We now validate our quantum orbit simulations with TDSE simulations. TDSE simulations are a more general approach and allow us to include tip-specific effects like the optical near-field decay present at the tip apex. Here, a 1D TDSE implementation with the single-active electron approximation is used~\cite{Dienstbier2023_nature, Dienstbier2023}, with a field enhancement factor of 5.5 at the tip surface and a decay length of 10\,nm for the optical near field (see Supplemental Material~\cite{SupplementalMaterial} for more details). The CEP map of the TDSE simulation is shown in Fig.~\ref{fig:Fig1_Colormaps}(b). In contrast to the quantum orbit result in Fig.~\ref{fig:Fig1_Colormaps}(a), the TDSE result also includes the direct electrons, which are dominant up to an energy of around $11 \, \mathrm{eV}$ and negligible in the plateau region at energies above $13 \, \mathrm{eV}$ in Fig.~\ref{fig:Fig1_Colormaps}(a). Apart from the direct electrons, we observe that the CEP maps look very similar in the plateau region. In particular, the arches of the intracycle interference maxima obtained from the quantum orbit simulation are overlaid in Fig.~\ref{fig:Fig1_Colormaps}(b), matching the TDSE low-frequency oscillations perfectly (see Supplemental Material~\cite{SupplementalMaterial} for more details). The excellent agreement between the quantum orbit and the TDSE results justifies the interpretation of the TDSE spectra with our quantum orbit model.

\begin{figure}[h!]
\includegraphics[width=\columnwidth]{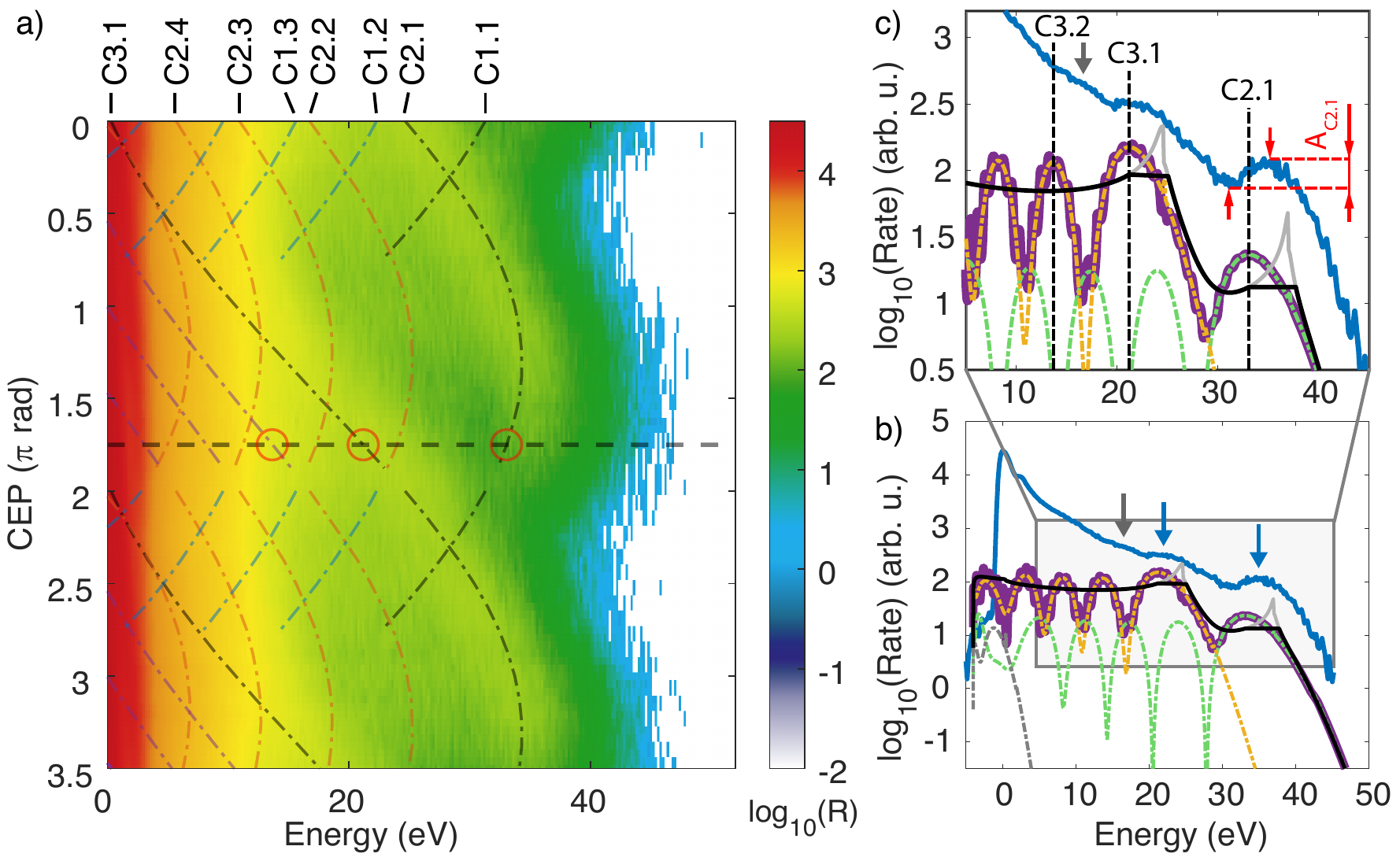}
\caption{\label{fig:Figure_3_240125_1459}Intracycle interference in experimentally measured photoemission spectra. (a) Experimental CEP map, with the energy resulting from the bias voltage of \SI{-30}{\volt} subtracted. The dash-dotted lines represent the intracycle interference maxima from the quantum orbit simulation and are overlaid to the experimental spectrum. Clearly, we can identify intracycle interference also in the experimental data: the overlaid lines match the experimental local maxima perfectly. Particularly clear are C2.1 and C3.1. We note that we applied an energy shift of 4.0\,eV to the quantum orbit results here and in the following to account for the optical near field of the tip; see text and Supplemental Material~\cite{SupplementalMaterial} for more details. (b) Lineout of (a): measurement (blue) and quantum orbit simulation (purple) for a CEP of $1.75 \, \pi \, \mathrm{rad}$ (dashed line in (a)). (c) Close-up of (b) in the plateau region. We can clearly see that the lineouts are mainly built up from the quantum orbit contributions C2 (green) and C3 (yellow). The appearance of C2.1 and C3.1 as peaks (blue arrows in (b)) is due to intracycle interference. At around 16\,eV, a small hump is just visible (gray arrow), which might hint at the intracycle inteference maximum C3.2. The quantum orbit expectations for the relevant intracycle interference maxima are highlighted with red circles in (a) and dashed lines in (c). In (b),(c) the incoherent addition of all contributions is shown in black, with artifact peaks in light gray and replaced by artificial horizontal lines. Clearly, the incoherent addition cannot explain the experimental data. Further lineouts of (a) are shown in Fig.~\ref{fig:Symbolic}(b). We note that the purple quantum orbit curve is smoothed here and thus does not show photon order peaks.}
\end{figure}

\section*{Intracycle interference in the measurement}

In Fig.~\ref{fig:Figure_3_240125_1459}(a), we overlay the quantum orbit results on the experimental data. There, the maxima C2.1 and C3.1 of the pair contributions from laser cycles 2 and 3 are clearly visible and perfectly match the energy-shifted quantum orbit values (dash-dotted black lines). The energy shift of the quantum orbit result is necessary to account for the optical near field (see Supplemental Material~\cite{SupplementalMaterial} for more details). The experimental spectra can now be directly explained by the energy-shifted quantum orbit results.

This is exemplary shown in Figs.~\ref{fig:Figure_3_240125_1459}(b) and~\ref{fig:Figure_3_240125_1459}(c) for a CEP of $1.75 \, \pi \, \mathrm{rad}$. In this case, the spectrum builds up from the contributions C2 and C3 with ionization times $t_0$ in the second (c2) and third laser cycle (c3). Figures~\ref{fig:Figure_3_240125_1459}(b) and~\ref{fig:Figure_3_240125_1459}(c) show two pronounced peaks in the measured spectrum at energies of around 22 and 34\,eV (blue arrows in (b)). Both peaks are reproduced by the quantum orbit simulation (purple curve) and originate from the intracycle interference maxima C2.1 and C3.1 at the cutoffs from both pair contributions. Without intracycle interference, the spectrum would appear steplike and without peaks, only indicating the classical cutoffs (black line in Figs.~\ref{fig:Figure_3_240125_1459}(b) and~\ref{fig:Figure_3_240125_1459}(c)). The signatures of the peaks at C2.1 and C3.1 can be further traced on the CEP map in Fig.~\ref{fig:Figure_3_240125_1459}(a), indicating the presence of intracycle interference at C2.1 and C3.1 for all CEPs. Apart from the clear peaks at C3.1 and C2.1, Figs.~\ref{fig:Figure_3_240125_1459}(b),(c) suggest that there is also a faint signal around 16\,eV (gray arrow) that roughly matches the expectation of the energy position of the intracycle interference maximum C3.2. Further discussion of the experimental data can be found in the Supplemental Material~\cite{SupplementalMaterial} in terms of a difference map, where the CEP average is subtracted from the CEP-resolved spectra of Fig.~\ref{fig:Figure_3_240125_1459}(a).

\section*{Electronic decoherence in the metal}

Based on these results, one might wonder why the intracycle interference is only pronounced at the highest energies. The answer could be electronic decoherence in the metal. The travel time difference between short and long orbits of a saddle-point pair increases toward lower energies (see Fig.~\ref{fig:Figure_2_QOVeranschaulichung_EM} in the End Matter and Supplemental Material~\cite{SupplementalMaterial}) and, consequently, decoherence processes become increasingly important, possibly washing out electron interference. In our case, the travel time difference is 1.1\,fs for the electrons between C[X].1 and C[X].2 (see Fig.~\ref{fig:Figure_2_QOVeranschaulichung_EM}(d)), but 1.4\,fs for C[X].2 and 1.8\,fs for C[X].3. In metals, electrons can lose coherence due to electron-electron scattering and electron-phonon scattering during driving~\cite{Rethfeld2002, Yanagisawa2011, Wendelen2013, Yanagisawa2016, Krueger2018, Heide2022}. Cooling of the metal tip or the choice of another material could lead to longer coherence times, and thus to a more pronounced intracycle interference pattern. Consequently, intracycle interference can be used to monitor and investigate the electronic coherence of solid materials in the form of nanometric needle tips. We extend this discussion in the End Matter. There, we also comment on the absence of clear intracycle interference signals in strong-field spectra from atoms in the gas phase and the role of focal averaging.

In conclusion, we have observed intracycle interference in photoemission spectra from a nanometric tungsten needle tip illuminated with two-cycle pulses at 1570\,nm central wavelength. We could clearly identify the intracycle interference by applying quantum orbit theory. To the best of our knowledge, this is the first clear experimental observation of intracycle interference of plateau electrons in any system so far. The intracycle interference leads to peaks close to the cutoffs of the saddle-point pair contributions from which the spectrum builds up. The absence of clear electron interference signals at lower energies in the photoemission spectra might be due to electronic decoherence in the metal tip. Stronger interference signals in photoemission spectra may be observed by increasing the coherence of the electronic ground state by cooling of the tip or by using a different tip material. Vice versa, our work now allows studying electron coherence on the 100-attosecond timescale without the need for attosecond light pulses. This is because the rescattering plateau appears highly pronounced, in fact even more pronounced than in spectra from atoms in the gas phase. In addition, the absence of focal averaging and the many degrees of freedom (tip size and material, application of static bias voltages, and the presence of optical near fields) make nanometric needle tips intriguing scattering targets for the investigation of complex electron interference effects in strong-field photoemission, on the attosecond timescale, with future insights to correlated materials. We also foresee deep insights by steering the electrons sideways and investigating the resulting spectra, in conjunction with a 2D quantum orbit analysis.

\begin{acknowledgments}
We thank Dejan B.\ Milošević and Wilhelm Becker for helpful discussions. This research was supported by the European Research Council (Advanced Grant AccelOnChip), the Gordon and Betty Moore Foundation (iQCE), and the DFG (Leibniz Prize). J.H.\ acknowledges funding from the Max Planck School of Photonics.
\end{acknowledgments}

\textit{Data availability}--The data that support the findings of this article are publicly available \cite{Wirth2026Data}.

\nocite{Thomas2013, Meier2023, Mattheis1965, Busuladzic2006, Rathje2012, Schmid2012, Thomas2015, Milosevic2024_topical, Habibovic2025, Milosevic2002_classification, Caldeira1983, Ingold2002, Weiss2008, Breuer2002}

\putbib[references.bib]

\clearpage\section*{End Matter}
\subsection*{Quantum orbit theory}

This section is an extension of the introduction in the main manuscript. Full details of the quantum orbit simulations used in this work can be found in the Supplemental Material~\cite{SupplementalMaterial}.
Quantum orbit theory~\cite{Lewenstein1995, Kopold2000, Kopold2001, Becker2002, Milosevic2006} is based on the strong-field approximation (SFA). More specifically, it relies on a saddle-point approximation of the respective SFA integral, where direct and rescattered electrons are treated separately. For this reason, the spectrum is calculated from a sum of contributions, which are individually determined by the solutions of the saddle-point equations, where the quantum mechanical action is stationary. These solutions are called saddle points and given by the triple $\left[t_0\left(\mathbf{p}\right), t_1\left(\mathbf{p}\right), \mathbf{k}\left(\mathbf{p}\right)\right]$, as outlined in the main text. Two saddle-point solutions belong together and represent short and long orbits.

\begin{figure}[h!]
\includegraphics[width=\columnwidth]{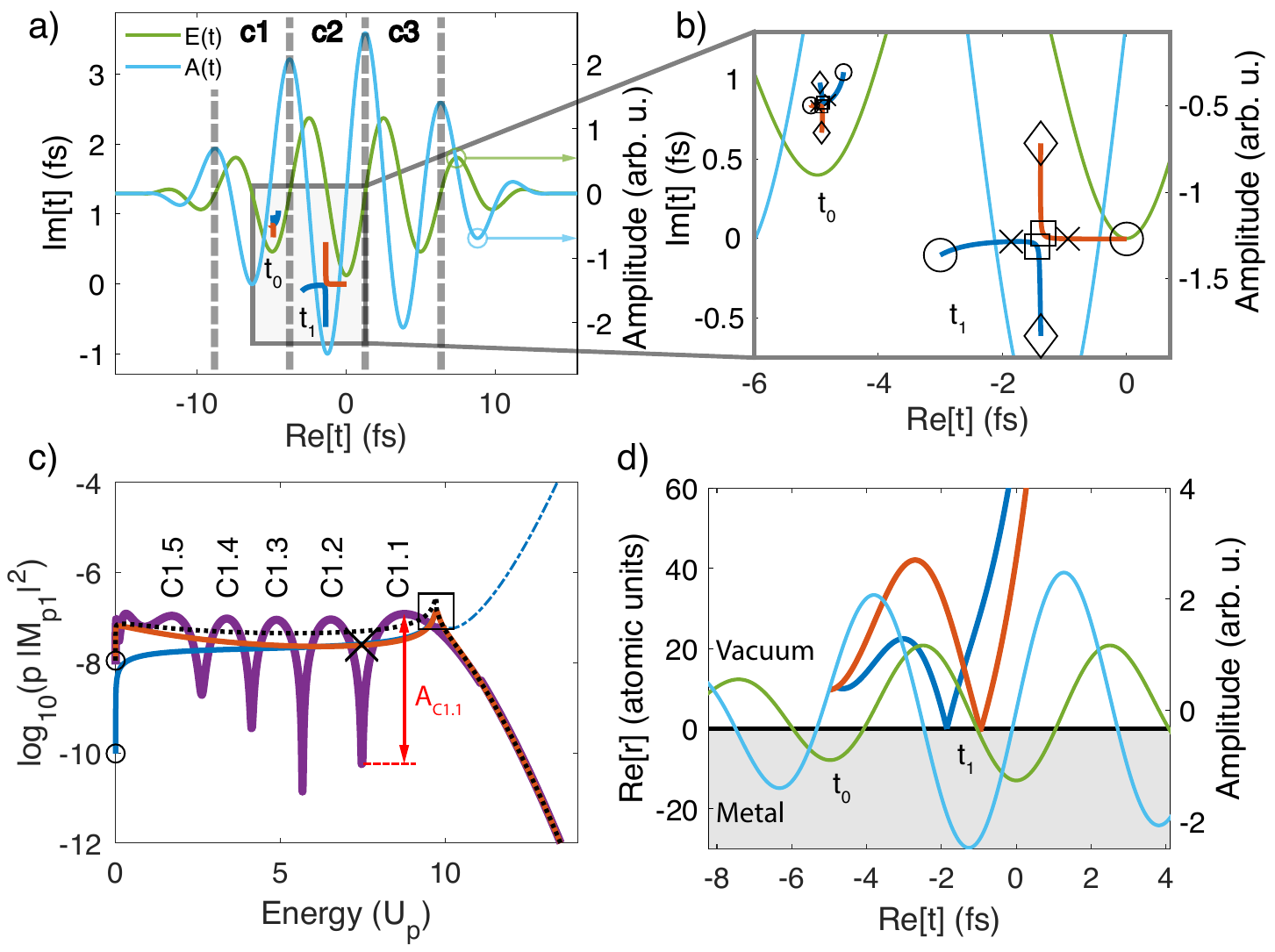}
\caption{\label{fig:Figure_2_QOVeranschaulichung_EM}Quantum orbit theory and the origin of intracycle interference. (a),(b): Two saddle-point solutions of the saddle-point pair C1 (ionization in laser cycle 1 (c1)) for various final energies $E_{\mathrm{f}}$ (lowest $E_{\mathrm{f}}$: circles, $E_{\mathrm{f}} = 7.45 \, U_{\mathrm{p}}$: cross, 10-$U_\mathrm{p}$~cutoff: squares, highest $E_{\mathrm{f}}$: diamonds), shown for $\mathrm{CEP} = 0$ and the parameters of the experiment. The associated optical electric field and vector potential are shown in green and blue. The saddle-point solutions constitute short (blue) and long (red) orbits. The time difference of emission for electrons reaching the 10-$U_\mathrm{p}$~cutoff is 23\,as, for example. (c) While the isolated spectral contributions $p {|M_{p1}|}^2$ of short and long orbits result in flat plateaus (dotted black line), their coherent sum (purple) shows a modulation with several humps. This modulation is due to interference of the two discussed trajectories, hence intracycle interference. There are similar contributions C2 and C3 with ionization times $t_0$ in laser cycle 2 (c2) and 3 (c3), from which the resulting full spectrum predominantly builds up. The intracycle interference maxima are labeled starting from the cutoff C1.1 toward lower energies. It should be noted that the sharp peaks at the $10$-$U_{\mathrm{p}}$~cutoff of the isolated contributions (red and blue lines) and their incoherent addition (dotted black line) are artificial and not present in their coherent addition (purple line), for which the uniform approximation was used. The trajectories of both orbits are shown in (d) for $E_{\mathrm{f}} = 7.45 \, U_{\mathrm{p}}$ (cross in (b),(c); energy of the dip between C1.1 and C1.2, indicated in (c) with the amplitude $A_{\mathrm{C1.1}}$). There, the time difference of emission for both electrons is 222\,as.}
\end{figure}

Figure~\ref{fig:Figure_2_QOVeranschaulichung_EM} is an extension of Fig.~\ref{fig:Figure_2_QOVeranschaulichung} in the main text. In addition to Fig.~\ref{fig:Figure_2_QOVeranschaulichung}, it contains a close-up view of the saddle points in the complex plane (Fig.~\ref{fig:Figure_2_QOVeranschaulichung_EM}(b)), and actual trajectories (Fig.~\ref{fig:Figure_2_QOVeranschaulichung_EM}(d)). The trajectories in Fig.~\ref{fig:Figure_2_QOVeranschaulichung_EM}(d) correspond to short (S, blue line) and long (L, red line) orbits for a final energy of $7.45 \, U_{\mathrm{p}}$ (30.1\,eV; energy at the minimum indicated by $A_{\mathrm{C1.1}}$ and corresponding to the cross in Figs.~\ref{fig:Figure_2_QOVeranschaulichung_EM}(b) and \ref{fig:Figure_2_QOVeranschaulichung_EM}~(c)). Here, the travel time difference between emission and rescattering for the short and long orbits amounts to $1.1 \, \mathrm{fs}$ (see crosses in Fig.~\ref{fig:Figure_2_QOVeranschaulichung_EM}(b) and $t_0$ and $t_1$ in Fig.~\ref{fig:Figure_2_QOVeranschaulichung_EM}(d)). The ionization $t_0$ and rescattering times $t_1$ are most different for short and long orbits at low final energies $E_{\mathrm{f}}$ (circles in Fig.~\ref{fig:Figure_2_QOVeranschaulichung_EM}(b) for the lowest $E_{\mathrm{f}}$) and will approach each other closely at the highest classically allowed final energy at the cutoff at $t_0 = \SI{-4.9}{fs}$ and $t_1 = \SI{-1.4}{fs}$ (squares). After the cutoff, the imaginary components of the saddle-point times $t_0$ and $t_1$ will increase (toward the diamonds for the highest $E_{\mathrm{f}}$), which in quantum orbit theory translates to a small spectral weight.

In Fig.~\ref{fig:Figure_2_QOVeranschaulichung_EM}(d), the metal half space is indicated in gray. For photoemission from metals, no trajectories are allowed in this half space. This metal boundary condition can be taken into account by selecting only those saddle-point solutions with trajectories that do not enter the metal surface. More details about this can be found in Supplemental Material~\cite{SupplementalMaterial}. Additionally, physically relevant saddle points must fulfill $\mathrm{Re}(t_1)>\mathrm{Re}(t_0)$ and $\mathrm{Im}(t_0)>0$.

We assumed laser pulses with a duration of 10\,fs and a central wavelength of 1550\,nm for all quantum orbit and TDSE simulations shown. Both parameters are well within the uncertainties of the experimental values, and especially the pulse duration of 10\,fs reproduces the CEP-dependent cutoff modulations in the measurement best.

\subsection*{Extented discussion: Electronic decoherence and the role of focal averaging}

As outlined in the main text, the amplitude of the intracycle interference pattern observed in the experiment is weaker than predicted by the fully coherent TDSE and quantum orbit simulations. This could be due to electronic decoherence in the tip material. Therefore, the intracycle interference in strong-field photoemission can be used to study the electronic coherence of any solid material in the form of a nanometric needle tip.

Nanometric needle tips can be obtained via chemical etching~\cite{Binh1996} or focused ion-beam milling~\cite{Thompson2007}. In addition, nanoscale samples can also be attached to the apex of a nanometric tip, which acts as a sample holder. The electronic coherence time inside a nanotip during a measurement can be estimated by directly comparing the measurement data with a suitable theoretical model that takes electronic decoherence into account. In Supplemental Material~\cite{SupplementalMaterial}, we introduce such a model~\cite{Heide2022} and discuss the effects of electronic decoherence in more detail. We note that electronic coherence can, in principle, also be investigated with high-harmonic generation in semiconductor materials~\cite{Heide2022}.

We emphasize that, although intracycle interference of rescattered electrons in the plateau region has not been clearly identified to date, faint signs of intracycle interference peaks close to the cutoff might be visible in~\cite{Paulus2003, Becker2018, Zhang2018}. In strong-field photoemission from atoms in the gas phase, the study of intracycle interference is obscured by focal averaging~\cite{Kopold2002CCs, PaulusGrasbon2003}. For example, a decrease in intensity of just $\sim$13\,\% shifts the intracycle interference maxima in the plateau region close to the destructive interference minima in Fig.~\ref{fig:Figure_3_240125_1459} (see Supplemental Material~\cite{SupplementalMaterial}), rendering intracycle interference highly sensitive to intensity variations such as in focal averaging. Closely related to this is the requirement for short pulses to resolve intracycle interference patterns. For longer pulses, intracycle interference contributions from additional laser cycles become relevant. Due to the slope of the laser pulse, these are related to different electric field strengths, and their intracycle interference patterns are shifted accordingly. In contrast to intracycle interference, typical low-energy structures (e.g.\ in~\cite{Khurelbaatar2024, Hasan2025}) change more slowly with intensity and are therefore less affected by focal averaging (see also~\cite{PaulusGrasbon2003}).

\end{bibunit}
\begin{bibunit}[apsrev4-2]

\clearpage
\onecolumngrid

\section*{Supplemental Material}

\setcounter{figure}{0}
\renewcommand{\figurename}{Fig.~S}

\section{Near-field-induced energy changes and CEP calibration of the measurement}\label{app:4eV}

Figure~3 of the main manuscript shows that the TDSE result is shifted by 4.0\,eV to lower energies compared to the quantum orbit result. This is most clearly visible at the high-energy maxima or corresponding arches and results from the optical near field present around the tip apex~\cite{Herink2012, Heimerl2025}. Optical near fields originate from electrons that are displaced at sharp metal features (dimension $r$) in the presence of a light field (wavelength $\lambda$), if $r \ll \lambda$~\cite{Thomas2013, Krueger2018, Dienstbier2023}.

Near-field effects cannot be directly incorporated into the quantum orbit simulation because of the strong-field approximation, which makes use of Volkov states. The effect of the near field depends on the specific electron trajectory. For a direct comparison of the quantum orbit result with the TDSE result, we tried to account for the near-field effects both by shifting and stretching the energy axis of the quantum orbit result. It is important to note that neither of the two adaptations fully accounts for near-field effects in the quantum orbit simulation. Here, we chose to shift the energy axis of the quantum orbit simulation for direct comparison with the TDSE and the experimental results. In order to experimentally determine the optical near-field-induced energy shift, we compared the CEP and energy positions of the maxima C2.1 and C3.1 in simulation and experiment.

In the experiment, CEP-resolved measurements were performed by subsequently changing the CEP in constant steps over roughly three optical periods with a pair of glass wedges. Naturally, absolute CEP values of the measurements are initially unknown. The CEP of the experiment was calibrated by the CEP of the quantum orbit simulation. First, the CEP periodicity of the highest cutoff energy was fitted with a sine, and the data were slightly interpolated in the CEP direction.

The experimental high-energy maxima C2.1 and C3.1 are obtained by fitting the maxima in the spectra, which are a function of the CEP and energy. Based on the quantum orbit simulation we calculated the mean squared error for different CEP and energy shifts to the experimental values, i.e., we created a 2D error map. The global minimum in this map then determined the CEP calibration of the measurement and, additionally, a near-field-induced energy shift of $-4.0 \, \mathrm{eV}$.

\section{Time-dependent Schrödinger equation (TDSE)}
\label{app:TDSE}

For our one-dimensional problem (light polarization and photoemission in the direction of the major axis of the tip), we use a one-dimensional TDSE implementation with a single-active electron (SAE approximation) in the length gauge~\cite{Dienstbier2023_nature, Dienstbier2023}. For propagation of the electron wavefunction, we use the Crank-Nicolson method.

More specifically, we use a Fermi-box binding potential that allows for one bound ground state and use a work function of $6.0 \, \mathrm{eV}$ for tungsten~\cite{Dienstbier2023_nature, Dienstbier2023, Meier2023}, a Fermi energy of $7.0 \, \mathrm{eV}$~\cite{Mattheis1965, Dienstbier2023}, a pulse with a Gaussian envelope, a pulse duration of $10 \, \mathrm{fs}$ (FWHM), and a peak near-field intensity of $I_{\mathrm{NF}} = 1.8 \cdot 10^{13} \, \mathrm{W/{cm}^2}$. The peak near-field intensity leads to a ponderomotive energy of $U_{\mathrm{p}} = 4.0 \, \mathrm{eV}$ and a Keldysh parameter of $\gamma = 0.9$. For the field enhancement factor, a value of 5.5, and for the decay length, a value of $10 \, \mathrm{nm}$ were chosen.

We note that the near field around the tip is spatially non-uniform, i.e., the near-field dynamic is three-dimensional. Still, we believe that our 1D-TDSE simulation sufficiently captures the relevant near-field dynamics, at least in our detection geometry. Our detector only picks up electrons from an opening angle of $9.2^\circ$ to the symmetry axis of the tip (see Fig.~S\ref{fig:fieldEnhancement}). At this angle, the optical near field is only 0.5\,\% smaller than that at the symmetry axis at the surface of the tip. We note that 1D-TDSE simulations were also used in~\cite{Heimerl2025} for the investigation of the NILES effect.

\section{Focal averaging and nanometric confinement at metal needle tips}\label{app:FocalAveraging}

The periodicity of the intracycle interference pattern of a saddle-point pair contribution is determined by the difference in the rescattering times between short and long orbits that interfere~\cite{Milosevic2006}. It is mainly fixed by the laser wavelength. However, the cutoff energy, roughly $10\,U_\mathrm{p}$ for rescattering at the laser pulse envelope maximum~\cite{Busuladzic2006}, is linearly proportional to the laser intensity $\left(U_\mathrm{p} = e^2 E_0^2/\left( 4 m \omega_\mathrm{opt}^2\right) \propto I\right)$. This makes intracycle interference signals prone to focal averaging~\cite{Kopold2002CCs, PaulusGrasbon2003}. Focal averaging may explain why, to the best of our knowledge, there are no reports of experimentally measured intracycle interference signals in strong-field photoemission spectra from atoms in the gas phase. There, the ATI process is happening for various intensities at different atoms inside the focal volume of the laser.

\begin{figure*}[h!]
\includegraphics[width=15cm]{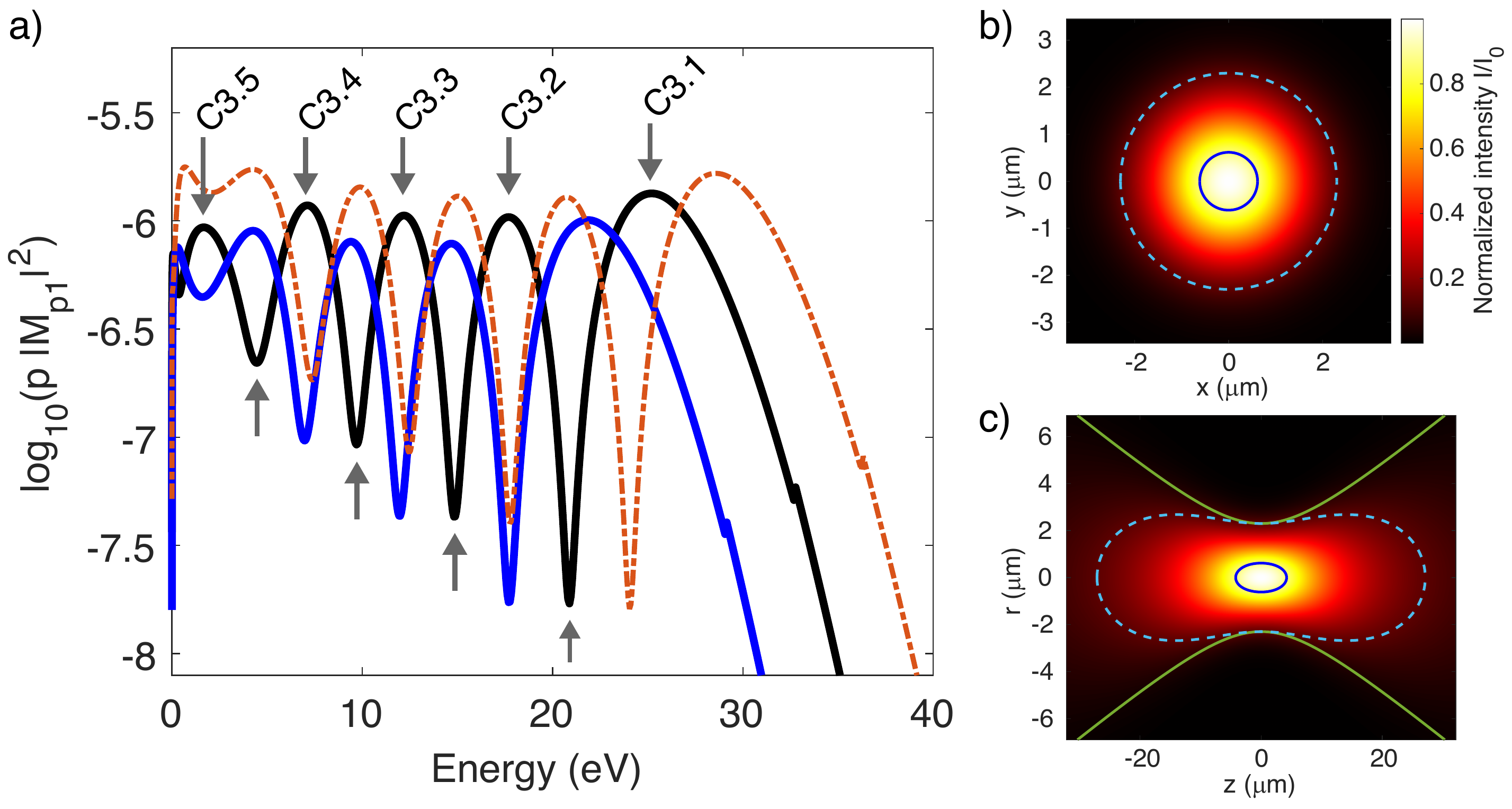}
\caption{\label{fig:FocAv}Focal averaging and nanometric confinement. (a) Simulated saddle-point pair contribution C3 at a CEP of $1.75 \, \pi \, \mathrm{rad}$ for the intensity of $18 \, \mathrm{TW/{cm}^2}$ (black line; intracycle interference maxima and minima are indicated with arrows) as in the main text, $15.6 \, \mathrm{TW/{cm}^2}$ (full blue line), and $20.4 \, \mathrm{TW/{cm}^2}$ (dash-dotted red line). The intracycle interference maxima of the blue and red spectra roughly shift to the minima of the black spectrum. This is the worst-case scenario for detecting intracycle interference. The intensities of the red and blue spectra differ by $\sim$13\,\% from $18 \, \mathrm{TW/{cm}^2}$, and are well covered by the intensity distribution of a Gaussian beam (b),(c). (b): (x,y)-plane, (c): (z,r)-plane of the normalized intensity distribution of a Gaussian beam ($\lambda=1550 \, \mathrm{nm}$ and a beam waist $w_0 = 2.3 \, \mathrm{\mu m}$ as in our experiment). The iso-intensity line at $0.867 \, I_0$ corresponds to the blue spectrum in (a) for $I_0 = 18 \, \mathrm{TW/{cm}^2}$ (black spectrum). The dashed cyan line in (b),(c) represents the iso-intensity line at $I_0(z)/e^2$ and the full green line in (c) the Gaussian beam waist, where the radial peak intensity $I_0(z)$ drops to $I_0(z)/e^2$. At the surface of the nanometric needle tip, the intensity is the near-field intensity $I_0$. Focal averaging is clearly not an issue for strong-field photoemission from nanometric needle tips. It should be noted that the intensity distribution around a metal nanotip is strongly affected by optical field enhancement and differs from a Gaussian beam profile (see text for more details).}
\end{figure*}

In Fig.~S\ref{fig:FocAv}(a), the saddle-point pair contribution C3 at a CEP of $1.75 \, \pi \, \mathrm{rad}$ (red circle in Fig.~4(a) in the main text) is shown for three intensities: black corresponds to a peak intensity of $I_0 = 18 \, \mathrm{TW/{cm}^2}$, as in the main text, blue to a peak intensity of $15.6 \, \mathrm{TW/{cm}^2}$, and red to $20.4 \, \mathrm{TW/{cm}^2}$. For the intensities of $15.6$ and $20.4 \, \mathrm{TW/{cm}^2}$, the intracycle interference maxima roughly shift to the minima of the black spectrum, which is the worst-case situation for the detection of intracycle interference. Both intensities differ only by $\sim$13\,\% from $I_0 = 18 \, \mathrm{TW/{cm}^2}$. To put this in context, in Figs.~S\ref{fig:FocAv}(b),(c) the intensity distribution of a diffraction-limited Gaussian beam with a beam waist of $w_0 = 2.3 \, \mathrm{\mu m}$ and $\lambda=1550 \, \mathrm{nm}$, as in our experiment, is shown in the (x,y)-plane at $z=0$ (beam waist plane) and in the (z,r)-plane $\left(r=\sqrt{x^2+y^2}\right)$. In (b) and (c), the iso-intensity line at $0.867\,I_0$ is shown (full dark line in the center), corresponding to the blue spectrum in (a). For comparison, we show the standard Gaussian beam waist (full green line in Fig.~S\ref{fig:FocAv}(c)). This illustrates that the intracycle interference maxima and minima from contributions of different intensities can easily average away. Consequently, intracycle interference may not be detectable in strong-field photoemission from atoms in the gas phase if the gas jet covers even only a fraction of the focal volume. Because gas jets typically exceed diameters of $100\, \mu m$~\cite{Rathje2012, Schmid2012}, focal averaging seems always relevant for focused laser light in the optical and near-infrared.

\begin{figure*}[h!]
\includegraphics[width=15cm]{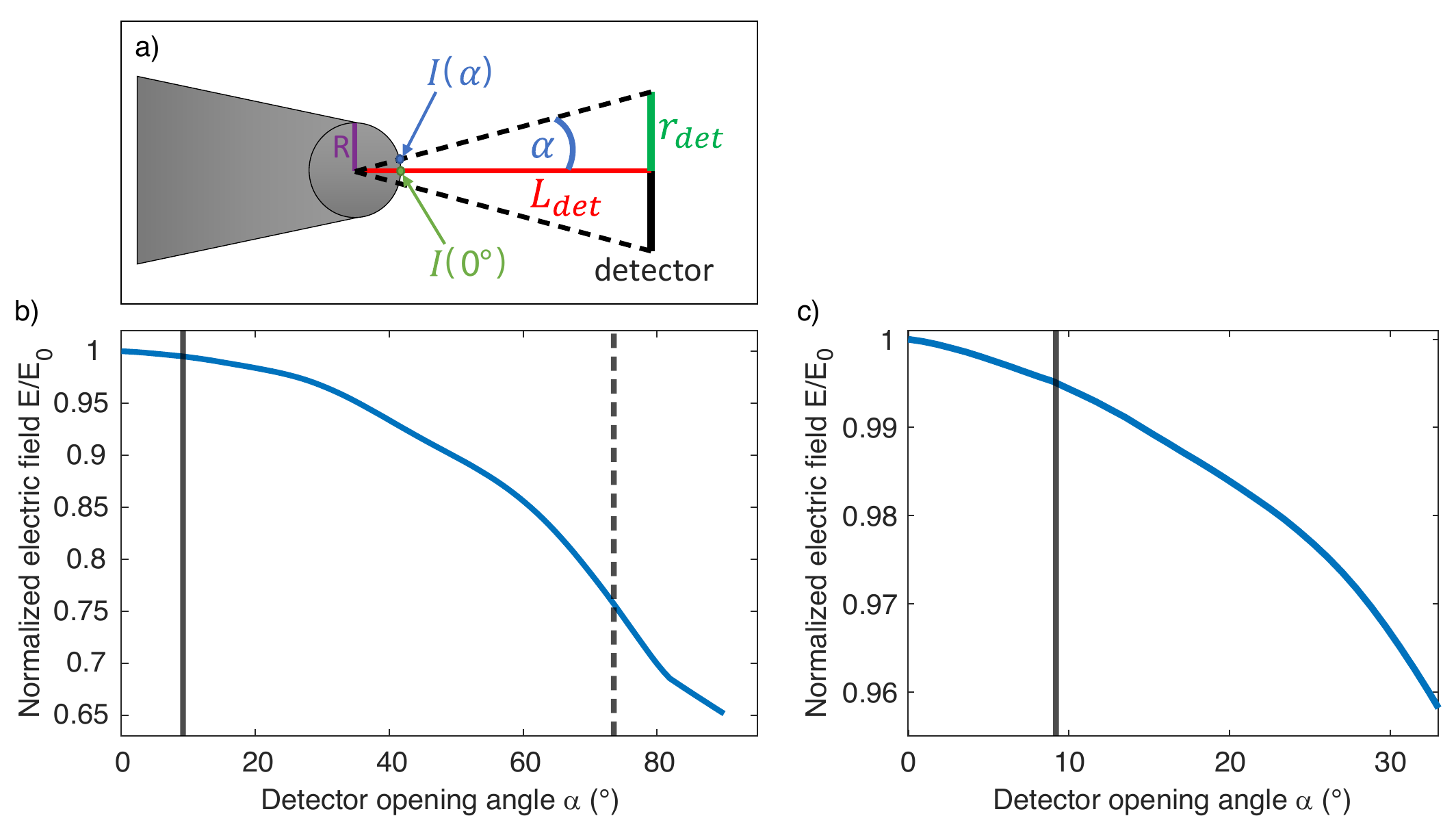}
\caption{\label{fig:fieldEnhancement}(a) Measurement geometry and the definition of the opening angle $\alpha$. (b) Numerical simulation results of the optical near-field decay on the surface of a nanometric tungsten needle tip with a radius of curvature $R$ of 10\,nm for illumination at a wavelength of 1600\,nm as a function of the detector opening angle $\alpha$. Angle dependence of the normalized optical near field on the tip surface. (c) Close-up of (b) for small opening angles. The opening angle of our measurement geometry is $9.2^\circ$ and indicated by the vertical full black line in (b) and (c). While the enhanced optical near field significantly decays for large opening angles, it only drops by $\sim$$0.5 \,\%$ at $\alpha = 9.2^\circ$ . This corresponds to an intensity drop of $\sim 1.0 \,\%$. This is well below the critical value of $\sim 13 \,\%$, which occurs at an opening angle of $\sim 74^\circ$ (vertical dashed line; see text for more details).}
\end{figure*}

In contrast, photoemission from nanometric needle tips is spatially confined to the tip apex, where optical field enhancement occurs. Because the tip apex of sharp needle tips like here lies in the range of 10\,nm, focal averaging is absent. As the cutoff energy is proportional to the intensity, the optical field enhancement effect, most pronounced at the apex of the needle, ensures that the plateau electrons originate from the apex of the tip. For tips with radii much smaller than the wavelength of the light, the field enhancement and therefore the near-field intensity are largest at the symmetry axis $\left(I_0\right)$ and decreases toward the shank~\cite{Herink2012, Thomas2013, Thomas2015, Krueger2018, Dienstbier2023, Heimerl2025}. Figure~S\ref{fig:fieldEnhancement} shows the optical field at and close to the surface of a nanometric tungsten needle tip with a curvature radius of 10\,nm and illumination at a wavelength of 1600\,nm, resulting from a numerical simulation~\cite{Thomas2015}.

Our measurement geometry is illustrated in Fig.~S\ref{fig:fieldEnhancement}(a). We used a detector with a radius $r_{\mathrm{det}}$ of 40\,mm placed $L_{\mathrm{det}}=247\,\mathrm{mm}$ away from the tip. This corresponds to a detector opening angle $\alpha$ of $9.2^\circ$ (full black line in Figs.~S\ref{fig:fieldEnhancement}(b),(c)). According to our near-field simulation in Fig.~S\ref{fig:fieldEnhancement}, the enhanced optical field drops by only $\sim$$0.5 \,\%$ if we go from on axis to an angle of $\alpha = 9.2^\circ$. This corresponds to an intensity drop of $\sim$$1.0 \,\%$. Based on the above discussion, this has a negligible effect on the position of the intracycle interference maxima positions. In particular, this is much less than the critical value of 13\,\%, where the intracycle interference maxima shift to the intracycle interference minima (see Fig.~S\ref{fig:FocAv}(a)). Hence, {\it angular} averaging is negligible in our measurement. In addition, we note that the detection of intracycle interference places high demands on the peak optical field uniformity (and stability) over the measurement volume or area.

\section{Quantum orbit simulation I: Theory and implementation}\label{app:QuantumOrbit}

This section provides more details of the quantum orbit simulation used in this work. The simulation is an implementation of the quantum orbit theory described in~\cite{Milosevic2006}. As outlined in the main manuscript, quantum orbit theory relies on the saddle-point approximation of strong-field integrals. Direct and rescattered electrons are treated separately. The strong-field approximation (SFA) of the rescattered electrons is also termed ISFA (improved SFA) to distinguish it from the SFA for the direct electrons (see, e.g.,~\cite{Milosevic2007CCs}). Similarly, the ATI process of the rescattered electrons is often termed "high-order above-threshold ionization" (HATI), whereas "above-threshold ionization" (ATI) is often used for direct electrons only~\cite{Milosevic2006}. Here, we use the term "ATI" to refer to both direct and rescattered electrons. Furthermore, we use "ATI" and "photoemission" interchangeably.

In this work, we are interested in the rescattered electrons, which lead to the formation of the plateau in the photoemission spectrum. To investigate this, we applied quantum orbit theory for the rescattered electrons. In the most general form, the strong-field integral of the rescattered electrons for a given final momentum $\mathbf{p}$ reads (atomic units are used throughout, unless stated otherwise)~\cite{Milosevic2006}:

\begin{equation}
\label{eq:Mp1}
M_{\mathbf{p}}^{\left(1\right)}=- \int_{t_{\mathrm{s}}}^{t_{\mathrm{f}}} \, dt_{1} \int_{t_{\mathrm{s}}}^{t_{1}} \, dt_{0} \int \, d^{3}\mathbf{k} \, V_{\mathbf{\tilde{p}}-\mathbf{k}} \braket{\mathbf{k}+\mathbf{A}\left(t_{0}\right)|\mathbf{r} \cdot \mathbf{E}\left(t_{0}\right)|\psi_{0}} \exp{\left[ i S\left( t_{0}, t_{1}, \mathbf{k}\right) \right]}
\end{equation}

$M_\mathbf{p}^{\left(1\right)}$ is a probability amplitude. The superscript "(1)" indicates rescattering (one interaction with the atom/tip; "(0)" is used for direct electrons). The detection probability for a given momentum $\mathbf{p}$ at the detector is given by $w\left(\mathbf{p}\right) = p |M_{\mathbf{p}}|^2$~\cite{Milosevic2006}. Here, $t_{\mathrm{s}}$ and $t_{\mathrm{f}}$ are the start and end times of the pulse, $t_0$ is the ionization time, $t_1$ is the rescattering time, and $\mathbf{k}$ is the drift momentum between $t_0$ and $t_1$. The dynamics of the process and the qualitative form of the photoemission spectrum are determined by the quantum mechanical action $S_{\mathbf{p}}$, for which

\begin{equation}
\label{eq:Wirkung}
S_{\mathbf{p}}\left(t_0, t_1, \mathbf{k} \right) = \int_{t_{\mathrm{s}}}^{t_1} d\tau \{{\left[ \mathbf{\tilde{p}} + \mathbf{A}\left(\tau\right)\right]}^2/2 + I_{\mathrm{p}}\} -\int_{t_0}^{t_1} d\tau \{{\left[ \mathbf{k} + \mathbf{A}\left(\tau\right)\right]}^2/2 + I_{\mathrm{p}}\}
\end{equation}

was used. $I_{\mathrm{p}}$ is the ionization potential of the atom or, respectively, the work function of the metal. Here, we use $I_{\mathrm{p}}>0$ throughout. In the matrix element $V_{\mathbf{\tilde{p}}-\mathbf{k}} = \langle \mathbf{\tilde{p}}| V | \mathbf{k} \rangle$, the interaction of the electron with the atom at the moment of rescattering is encoded, with $V$ denoting the rescattering potential. The use of $\mathbf{\tilde{p}} = \mathbf{p} - \mathbf{A}\left(t_{\mathrm{f}}\right)$ takes into account that the vector potential at the end of the pulse $t_{\mathrm{f}}$ can generally be non-zero and ensures that $M_{\mathbf{p}}^{\left(1\right)}$ is calculated for $\mathbf{p}$ at the detector if $\mathbf{A}\left(t_{\mathrm{f}}\right)\neq0$. In the experiment, the light pulses were linearly polarized, and the electrons were detected in the direction of light polarization in the positive half space (away from the metal surface). Therefore, a vector potential $\mathbf{A}\left(t\right) = \left(A_x, 0, 0\right)$ and an electric field $\mathbf{E}\left(t\right) = \left(E_x, 0, 0\right)$ with

\begin{equation}
\label{eq:Vektorpot}
A_x\left(t\right) = A_{0} {\cos}^2{\left(\frac{\omega t}{2 n}\right)} \sin{\left(\phi_{\mathrm{CEP}} + \omega t \right)}
\end{equation}

\begin{equation}
\label{eq:EFeld}
E_x\left(t\right) = -\frac{\partial A_x\left(t\right)}{\partial t} =-A_{0} \omega {\cos}^2{\left(\frac{\omega t}{2 n}\right)} \cos{\left(\phi_{\mathrm{CEP}} + \omega t \right)} + \frac{A_0 \omega}{n} \cos{\left(\frac{\omega t}{2 n}\right)} \sin{\left( \frac{\omega t}{ 2 n}\right)} \sin{\left( \phi_{\mathrm{CEP}}+\omega t\right)}
\end{equation}

and

\begin{equation}
\label{eq:n_fuer_Vektorpot}
n=\frac{\omega \, \tau_{\mathrm{FWHM}}}{4 \arccos{\left(2^{-1/4}\right)}}
\end{equation}

were used, similarly as described in~\cite{Milosevic2006}. In this case, $\mathbf{A}\left(t_{\mathrm{f}} \right) = 0$, and consequently, $\mathbf{\tilde{p}}$ can always be replaced by the momentum $\mathbf{p}$ at the detector. For this pulse definition, $t_{\mathrm{s}}$ and $t_{\mathrm{f}}$ are given by $-n \, \pi$ and $n \, \pi$ (see Eq.~\eqref{eq:n_fuer_Vektorpot}). The strong-field approximation makes use of Volkov states. Therefore, $|\mathbf{\tilde{p}}\rangle$ and $| \mathbf{k} \rangle$ are plane wave states, e.g.:

\begin{equation}
\label{plane wave explicit}
\langle \mathbf{r} | \mathbf{\tilde{p}} \rangle = \left(2 \, \pi\right)^{-3/2} \exp{\left(i \mathbf{\tilde{p}} \cdot \mathbf{r}\right)}
\end{equation}

The matrix element $\braket{\mathbf{k}+\mathbf{A}\left(t_{0}\right)|\mathbf{r} \cdot \mathbf{E}\left(t_{0}\right)|\psi_{0}}$ in the probability amplitude~\eqref{eq:Mp1} is called the ionization dipole matrix element (IDME)~\cite{Kuebel2021}. $\psi_{0} =\psi_{0}\left(\mathbf{r}\right)$ is the electronic ground state wave function, or, to be more precise, the spatial component of the time-dependent electronic ground state wave function $\psi_{0}\left(\mathbf{r}, t\right) = \exp{\left( i I_{\mathrm{p}} t\right)} \, \psi_{0}\left(\mathbf{r}\right)$. In Eq.~\eqref{eq:Mp1}, the time dependence of the electronic ground state $|\psi\left(\mathbf{r}, t\right)\rangle$ is already separated from the spatial integration and grouped into the action term (for more details see~\cite{Becker2002, Milosevic2006}). According to~\cite{Milosevic2006}, the ionization dipole matrix element can be simplified to

\begin{equation}
\label{eq:DME}
\braket{\mathbf{k}+\mathbf{A}\left(t_{0}\right)|\mathbf{r} \cdot \mathbf{E}\left(t_{0}\right)|\psi_{0}} = -i 2^{7/2} \left( 2 I_{\mathrm{p}}\right)^{5/4} \frac{\left[\mathbf{k} + \mathbf{A}\left( t_0\right)\right] \cdot \mathbf{E}\left(t_0\right)}{\pi \, \left(\left[\mathbf{k}+\mathbf{A}\left(t_0\right)\right]^2 + 2 I_{\mathrm{p}} \right)^3}
\end{equation}

for a hydrogen-like ground-state wave function $\psi_{0}\left(\mathbf{r}\right)$. For a more detailed investigation and a treatment of different ground-state wave functions, see, e.g.,~\cite{Kuebel2021} and~\cite{Milosevic2024_topical}. Here, Eq.~\eqref{eq:DME} was used to model photoemission spectra from a metal nanotip. The ionization dipole matrix element is discussed in more detail at the end of this section.

In the rescattering potential matrix element $V_{\mathbf{\tilde{p}}-\mathbf{k}} = \langle \mathbf{\tilde{p}}|V|\mathbf{k}\rangle$ we used a Yukawa-type potential for the rescattering potential $V\left(r\right)$, leading to

\begin{equation}
\label{eq:Vpk}
V_{\mathbf{\tilde{p}}^-\mathbf{k}} = - \frac{2 b \lambda + a C}{2 \, \pi^2 C^2}
\end{equation}

with

\begin{equation}
\label{eq:Vpk_C}
C = \left( \mathbf{\tilde{p}}-\mathbf{k} \right)^2 + \lambda^2
\end{equation}

and the parameters $a=1$, $b=0$, $\lambda = 1$, analogous to~\cite{Milosevic2006}. The effects of $V_{\mathbf{\tilde{p}}-\mathbf{k}}$ and the ionization dipole matrix elements on the appearance of the ATI spectrum are discussed in more detail at the end of this section. The influence of the atomic structure on the appearance of ATI spectra was investigated in~\cite{Habibovic2025}.

Quantum orbit theory relies on a saddle-point approximation of the respective strong-field integral, which means Eq.~\eqref{eq:Mp1} for the rescattered electrons. For this, first, the integration over the components of the intermediate drift momentum $\mathbf{k}$ in Eq.~\eqref{eq:Mp1} is eliminated, taking into account Eq.~\eqref{eq:DME}. This yields (for details see~\cite{Milosevic2006}):

\begin{equation}
\label{eq:M2006_B6}
M_{\mathbf{p}}^{\left(1\right)} =  \int_{t_{\mathrm{s}}}^{t_{\mathrm{f}}} \, dt_1 \int_{t_{\mathrm{s}}}^{t_1} \ dt_0 F {S'}^{-1} \exp{\left( i S\right)}+\mathrm{bt}
\end{equation}

with "bt" being a boundary term that is not relevant for quantum orbit theory, and the equations:

\begin{equation}
\label{eq:M2006_B7}
F = \frac{2^{7/2} \left( 2 I_{\mathrm{p}}\right)^{5/4}}{16 \, \pi} V_{\mathbf{\tilde{p}}-\mathbf{k_s}} \left[ \frac{2 \, \pi}{i \left(t_1 - t_0 \right)}\right]^{3/2}
\end{equation}

\begin{equation}
\label{eq:M2006_S_prime}
S' = \frac{\partial S}{\partial t_0 } = \frac{1}{2} \left[ \mathbf{k_{s}} + \mathbf{A}\left( t_0\right) \right]^2 + I_p
\end{equation}

\begin{equation}
\label{eq:ks}
\mathbf{k_s} = - \int_{t_0}^{t_1} dt \mathbf{A}\left(t\right)/\left( t_1 - t_0\right)
\end{equation}

Due to the reduced dimensionality, Eq.~\eqref{eq:M2006_B6} is also particularly advantageous for numerical integration. It should be noted that the exponent $3/2$ in Eq.~\eqref{eq:M2006_B7} results from three times $1/2$ for each spatial dimension. For a one-dimensional simulation, this factor would be $1/2$. Here, we calculated the probability amplitude in one direction using the 3D~simulation, as was also done in~\cite{Kuebel2021}. Starting from Eq.~\eqref{eq:M2006_B6}, the saddle-point approximation of $M_{\mathbf{p}}^{\left(1\right)}$, according to~\cite{Milosevic2006}, is given by

\begin{equation}
\label{eq:M2006_B10}
M_{\mathbf{p}}^{\left( 1 \right)} = i \, \pi \sum_{s} F\left( t_{1s}, t_{0s}\right) \left( \frac{\partial^2 S_s}{\partial t_0^2}\right)^{-1} \left( \frac{2 \, \pi \, i}{\partial^2S_s/\partial t_1^2}\right)^{1/2} \exp{\left( iS_s\right)}
\end{equation}

with

\begin{equation}
\label{eq:M2006_B10_Teil1}
\frac{\partial^2 S_s}{\partial t_0^2} =  \frac{{\left[ \mathbf{k_s} + \mathbf{A}\left(t_0\right) \right]}^2}{t_1 - t_0} -\left[\mathbf{k_s}+\mathbf{A}\left(t_0\right)\right] \cdot \mathbf{E}\left(t_0\right)
\end{equation}

and

\begin{equation}
\label{eq:M2006_B10_Teil2}
\partial^2S_s/\partial t_1^2 =\left(\mathbf{k_s} - \mathbf{p}\right) \cdot \mathbf{E}\left(t_1\right) + \frac{1}{t_1-t_0}\left[ \left[ \mathbf{k_s} + \mathbf{A}\left(t_1\right)\right]^2 - \frac{\left( \left[ \mathbf{k_s}+\mathbf{A}\left(t_0\right)\right] \cdot \left[\mathbf{k_s}+\mathbf{A}\left(t_1\right) \right]\right)^2}{\left[ \mathbf{k_s}+\mathbf{A}\left(t_0\right)\right]^2 - \left[ \mathbf{k_s}+\mathbf{A}\left(t_0\right)\right] \cdot \mathbf{E}\left(t_0\right) \left( t_1-t_0\right)}\right]
\end{equation}

The subscript $s$ denotes the sum over all saddle-point solutions. These are the solutions of the saddle-point equations. The saddle-point equations are determined by the condition that the quantum mechanical action with respect to $t_0$, $t_1$, and the components of $\mathbf{k}$ $\left(k_i\right)$ is stationary, i.e., the condition that their partial derivatives vanish. The saddle-point equations are given by Eq.~\eqref{eq:SPT2} $\left( \partial S/\partial k_i = 0\right)$, Eq.~\eqref{eq:SPT_t0} $\left( \partial S/\partial t_0 = 0\right)$, and Eq.~\eqref{eq:SPT_t1} $\left( \partial S/\partial t_1 = 0\right)$. In practice, determining the saddle points means solving equations \eqref{eq:SPT_t0} and \eqref{eq:SPT_t1} with $\mathbf{k_s}\left(t_0, t_1\right)$ in the form of Eq.~\eqref{eq:ks} (drift momentum $\mathbf{k_s}$ is constant for a given set of $t_0$ and $t_1$) inserted.

\begin{equation}
\label{eq:SPT2}
\int_{t_0}^{t_1} d\tau \left[ \mathbf{k_s} + \mathbf{A}\left(\tau\right)\right]=0
\end{equation}

\begin{equation}
\label{eq:SPT_t0} 
\left[ \mathbf{k_s} + \mathbf{A}\left(t_{0}\right) \right]^2 = -2 I_{\mathrm{p}}
\end{equation}

\begin{equation}
\label{eq:SPT_t1} 
\left[ \mathbf{k_s} + \mathbf{A}\left(t_{1}\right) \right]^2 = \left[ \mathbf{\tilde{p}} + \mathbf{A}\left(t_{1}\right) \right]^2
\end{equation}

\begin{figure*}[h!]
\includegraphics[width=9cm]{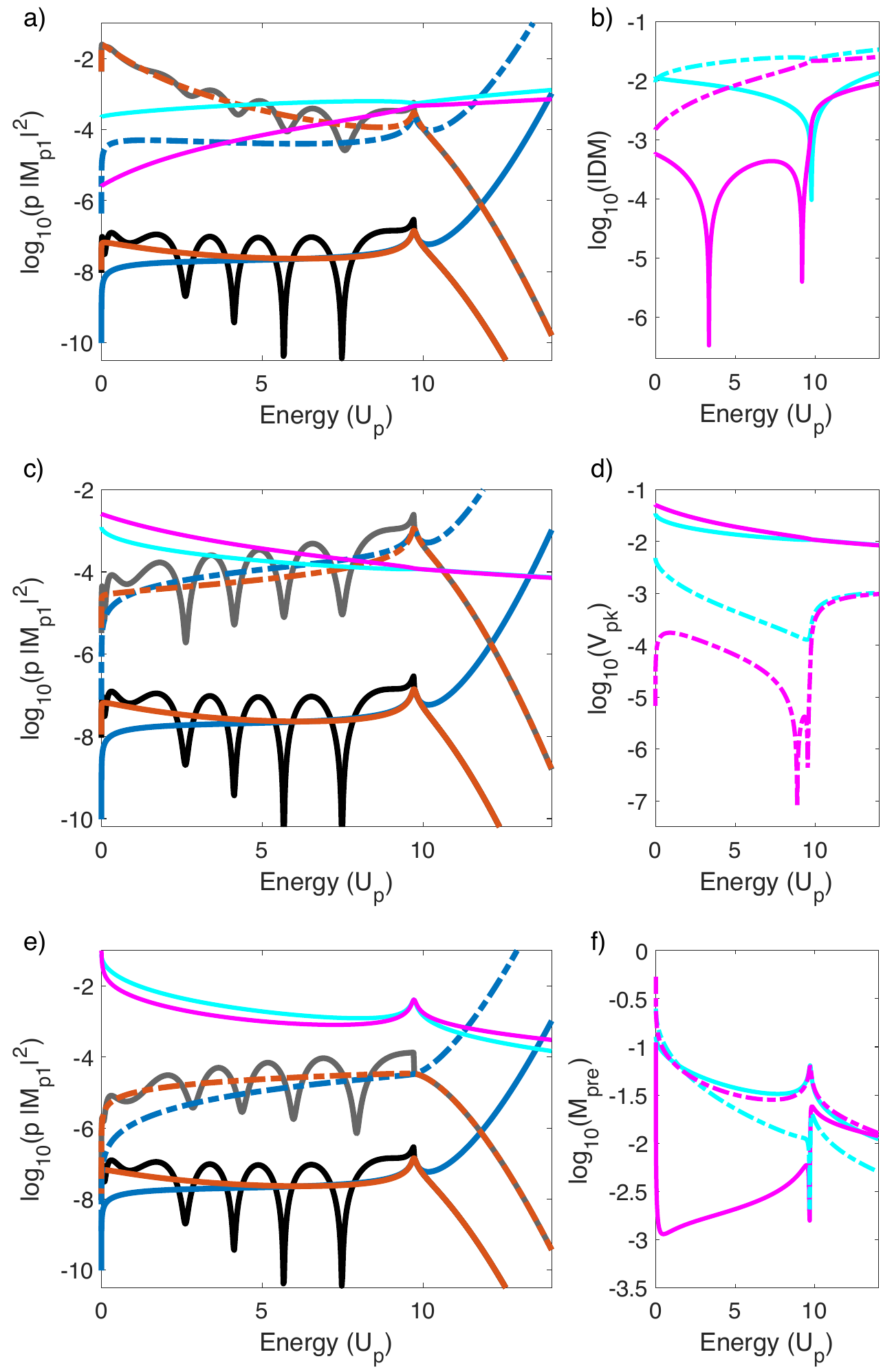}
\caption{\label{fig:Supp_IDM}Influence of IDME (a,b), $V_{\mathbf{\tilde{p}}-\mathbf{k}}$ (c,d), and the rest term $M_{\mathrm{rest}}$ ($M_{p1} = M_{\mathrm{rest}} \exp{\left(iS\right)}$) (e,f) on the spectrum in our quantum orbit simulation, exemplarily shown for C1 at $\mathrm{CEP} = 0$. Blue and red full lines in the left panels represent the full spectral yields $p |M_{p1}|^2$ of short (blue) and long (red) orbits (coherent sum in black; like in Fig.~2 in the main manuscript), while the dash-dotted lines show the same spectral yields but without taking into account IDME (a), $V_{\mathbf{\tilde{p}}-\mathbf{k}}$ (c) and $M_{\mathrm{rest}}$ (e) ($M_{p1} = M_{\mathrm{rest}} \exp{\left(iS\right)}$), respectively (coherent sum in gray). $|\mathrm{IDME}|^2$ (a), $|V_{\mathbf{\tilde{p}}-\mathbf{k}}|^2$ (c), and $|M_{\mathrm{rest}}|^2$ (e) (cyan for short and pink for long orbits) are spectrally flat on logarithmic scale, especially in the plateau region from $4 \, U_\mathrm{p}$ to $10 \, U_\mathrm{p}$, and do not have significant influence on the energy positions of the intracycle interference maxima and minima (black and gray lines in (a) and (c)). Real (full line) and imaginary parts (dash-dotted lines) of IDME (b), $V_{\mathbf{\tilde{p}}-\mathbf{k}}$ (c), and $M_{\mathrm{rest}}$ (d) are shown in the right panels. While for IDME and $V_{\mathbf{\tilde{p}}-\mathbf{k}}$ either the imaginary parts or the real parts dominate, the situation is more complex for $M_{\mathrm{rest}}$ (f). As a consequence, the energy positions of the intracycle interference maxima and minima are not correctly described by only taking into account $\exp{\left(iS\right)}$ for the spectral yields (black and gray lines in (e).}
\end{figure*}

Equations \eqref{eq:SPT_t0} and \eqref{eq:SPT_t1} state that energy $\left[E=p^2/\left(2 m \right)\right]$ is conserved at the times of ionization $t_0$ and rescattering $t_1$, while Eq.~\eqref{eq:SPT2} states that the rescattering process occurs at the same place where the electron was born $\left[ \int d\tau \,\mathbf{v}=0\right]$. One solution of the saddle-point equations is exemplarily shown in Fig.~2 in the main text. The buildup of the spectrum in quantum orbit theory is discussed in more detail below. Here, we want to discuss the effects of the ionization dipole matrix element (IDME) and the rescattering potential matrix element $V_{\mathbf{\tilde{p}}-\mathbf{k}}$ in our simulations. In Figs.~S\ref{fig:Supp_IDM}(a),(c) we show the spectral contribution of C1 at CEP=0 (same as in Fig.~2 in the main text), with and without IDME (a) and $V_{\mathbf{\tilde{p}}-\mathbf{k}}$ (b), together with $|\mathrm{IDME}|^2$ and $|V_{\mathbf{\tilde{p}}-\mathbf{k}}|^2$ for direct comparison.

In the left-hand panels of Fig.~\ref{fig:Supp_IDM}, we plot the isolated spectral contributions of short (full blue lines) and long (full red lines) orbits, together with their coherent sum (full black lines), as shown in Fig.~2 in the main text. To evaluate the effects of IDME, $V_{\mathbf{\tilde{p}}-\mathbf{k}}$, and $M_{\mathrm{rest}}$ ($M_{p1} = M_{\mathrm{rest}} \exp{\left(iS\right)}$) on the spectrum, we show their absolute squares in cyan (short orbit) and pink (long orbit) and the spectral yield without them (dash-dotted lines for the isolated spectral contributions and full gray line for their coherent sum). As we can see from Fig.~S\ref{fig:Supp_IDM}, $|\mathrm{IDME}|^2$ (a) and $|V_{\mathbf{\tilde{p}}-\mathbf{k}}|^2$ (c) are spectrally flat on logarithmic scale and do not significantly change the form of the spectrum, especially in the plateau region at energies above $5 \, U_{p}$. Importantly, the energy positions of the maxima and minima of the intracycle interference pattern with (black) and without $|\mathrm{IDME}|^2$ (a) and $|V_{\mathbf{\tilde{p}}-\mathbf{k}}|^2$ (c) (gray) are almost identical. They are not significantly affected by IDME and $V_{\mathbf{\tilde{p}}-\mathbf{k}}$.

In the plateau region, both matrix elements only shift the spectra on the ordinate to good approximation. We can think of them in a simplified way as global factors for the detection probabilities in a certain direction of the final momentum $\mathbf{p}$. In this work, measurements and simulations were only performed for one direction of the final electron momentum $\mathbf{p}$, and the simulation results (quantum orbit and TDSE simulations) were globally shifted on the ordinate for direct comparison. Therefore, the use of atomic parameters in IDME and $V_{\mathbf{\tilde{p}}-\mathbf{k}}$ in our simulations is justified for comparison and interpretation with our experimental data. We do not expect any significant changes in the form of quantum orbit spectra at energies above $1\,U_\mathrm{p}$, except shifts on the ordinate when choosing a different ground-state wave function $\psi_0\left(r\right)$ (e.g., that of a step potential) in the IDME or a different rescattering potential in $V\left(r\right)$.

We note that, although IDME and $V_{\mathbf{\tilde{p}}-\mathbf{k}}$ do not lead to significant changes in the form of the intracycle interference pattern, it is important to include the prefactors of the saddle-point approximation in the simulation results. Fig.~S\ref{fig:Supp_IDM}(e) shows the spectral yield of only the exponential term $\exp{\left(iS\right)}$ and the absolute square of the rest term $M_{\mathrm{rest}}$ ($M_{p1} = M_{\mathrm{rest}} \exp{\left(iS\right)}$) in the same way as with IDME and $V_{\mathbf{\tilde{p}}-\mathbf{k}}$. Here, the energy positions of the maxima and minima of the black and gray lines are shifted. Hence, it is important to include the prefactors of the saddle-point approximation in the simulation results to correctly describe the intracycle interference pattern. The combination of IDME and $V_{\mathbf{\tilde{p}}-\mathbf{k}}$, i.e., $\mathrm{IDME} \, \cdot \, V_{\mathbf{\tilde{p}}-\mathbf{k}}$ as a prefactor of the spectral yield, does not lead to significant energy shifts of the intracycle interference pattern, similarly to (a) and (e). We note that the intracycle interference pattern is determined by the electron dynamics and not the spatial component of the ground-state wave function $\psi_0\left(r\right)$ and the rescattering potential $V\left(r\right)$, which are not time dependent.

For the sake of completeness, in the right-hand panels of Fig.~S\ref{fig:Supp_IDM}, we show the real parts of IDME (b), $V_{\mathbf{\tilde{p}}-\mathbf{k}}$ (d), and $M_{\mathrm{rest}}$ (f) in full lines and their imaginary parts with dash-dotted lines. The presence of imaginary components is directly linked to the saddle points, which are complex quantities. We find that for IDME and $V_{\mathbf{\tilde{p}}-\mathbf{k}}$ either the imaginary parts (IDME) or the real parts $V_{\mathbf{\tilde{p}}-\mathbf{k}}$ are dominant, while for the rest term $M_{\mathrm{rest}}$ both parts are important.

\section{Quantum orbit simulation II: Buildup and interpretation of photoemission spectra}\label{app:KohInkoh}

In quantum orbit theory, the spectrum is calculated from the coherent sum $p |\sum_{i=1} M_i|^2$ of the spectral weights of all physically relevant saddle-point solutions (with $\mathrm{Re}(t_1)>\mathrm{Re}(t_0)$ and $\mathrm{Im}(t_0)>0$). In contrast to photoemission from atoms, electrons with trajectories in the negative half space (inside the metal) cannot contribute to the photoemission spectrum. These electrons are most likely lost inside the metal. This is because the electric field is strongly screened inside the metal. Thus, there is no (substantial) electric field in the negative half space that can act on these electrons. They will eventually be scattered or absorbed inside the tip and will not contribute to the photoemission spectrum.

\begin{figure*}[h!]
\includegraphics[width=15cm]{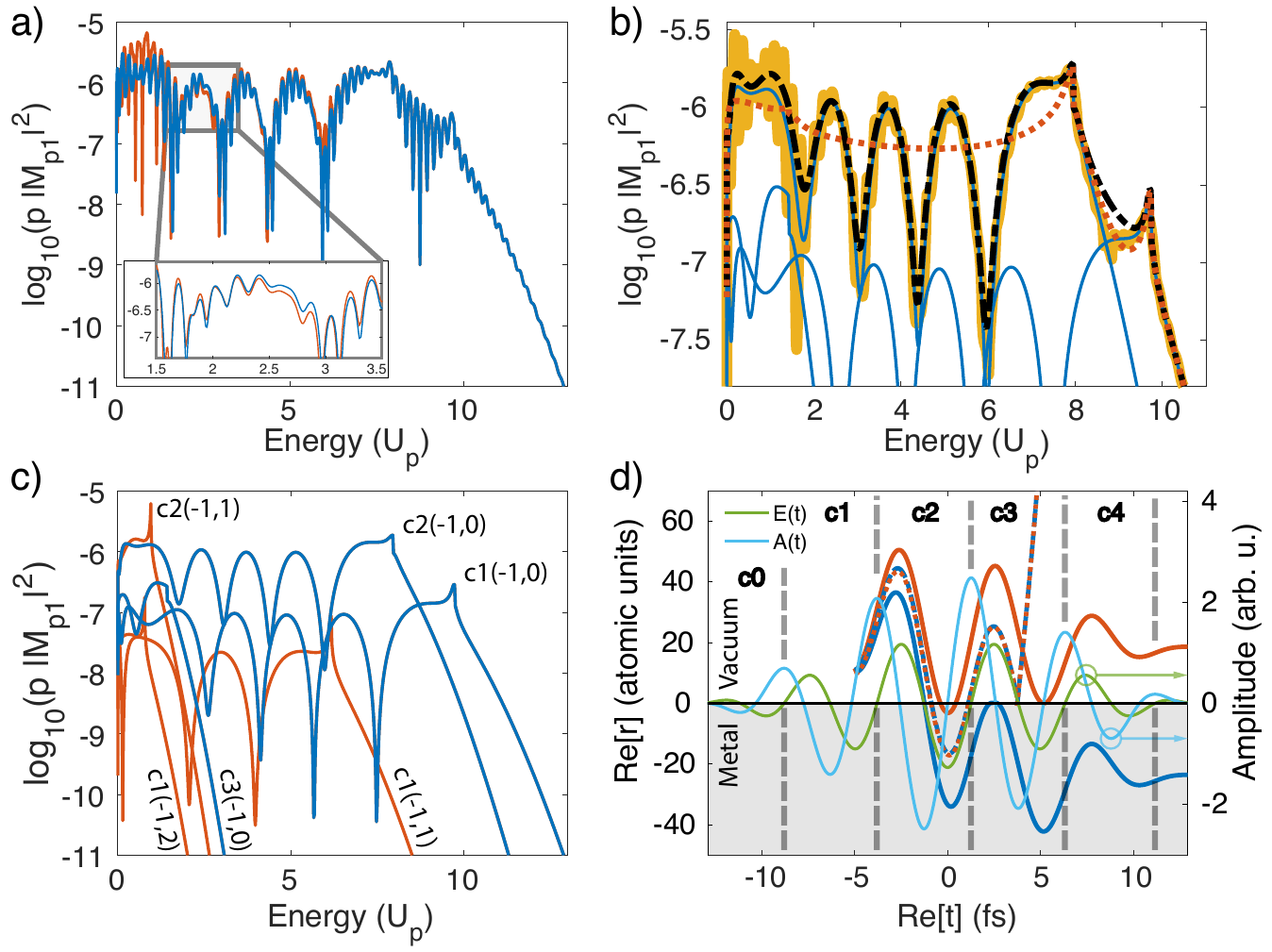}
\caption{\label{fig:alleTraj}Short and long pair contributions and explanation of photoemission spectra. (a) Quantum orbit simulations for the parameters of the experiment and $\mathrm{CEP}=0$ in which only the short pair contributions c1($-1$,0), c2($-1$,0), and c3($-1$,0) are considered (blue), and additionally the longer contributions are taken into account (red). These contributions are shown in (c). The red pair contributions have either low cutoff energies or low spectral weights and have negligible influence on the simulation result. In the plateau region ($E_{\mathrm{f}}>2\,U_{\mathrm{p}}$), only c1($-1$,1) leads to small differences between the red and the blue curves in a. In (d), electron trajectories of c1($-1$,1) (red: long and blue: short orbit) are shown for $E_{\mathrm{f}}=1.5 \cdot 10^{-5}\,U_{\mathrm{p}}$ (solid lines) and $14.9\,U_{\mathrm{p}}$ (dotted red and dash-dotted blue lines), well above the 10-$U_{\mathrm{p}}$ cutoff. The red and blue trajectories are most different for the lowest $E_{\mathrm{f}}$ and come close at the cutoff and at energies above. The trajectories of all electrons of c1($-1$,1) lead into the negative half space (inside the metal) and should be excluded. In (b), the blue photoemission spectrum of a is shown in yellow together with the short pair contributions, from which it is constructed. The yellow line has been smoothed to eliminate distracting photon order peaks. In b, the incoherent addition of the blue pair contributions is shown by the black dash-dotted line. It closely matches the yellow line, which is typical for all CEPs. For this reason, the photoemission spectra can be easily explained by just plotting the pair contributions. In (a--c), no uniform approximation was used. Therefore, artificial spikes are visible at the cutoff energies of the pair contributions.}
\end{figure*}

Figure~S\ref{fig:alleTraj}(a) shows the photoemission spectra of two quantum orbit simulations for the parameters of the experiment and $\mathrm{CEP}=0$. In the blue spectrum, only the shortest pair contributions of the three main cycles (C1, C2, and C3) are taken into account, as has been done in the main manuscript for this CEP. They are shown in Fig.~S\ref{fig:alleTraj}(c) in blue and are labeled according to the main cycles in which ionization occurs and ($\beta$,\,$m$) in the classification scheme of~\cite{Milosevic2002_classification}. In this notation, $m$ roughly indicates the travel time between ionization and rescattering in units of optical cycles, with $m=0$ denoting the shortest travel times. $\beta=-1$ if ionization occurs at negative electric field and $\beta=1$ if ionization occurs at positive electric field. In the case of ionization at negative electric field, the electron is born in the positive half space at a certain tunnel exit, depending on the electric field strength, and with a trajectory away from the metal surface. In contrast, in the case of ionization at positive electric field, the electron is born in the negative half space at a certain tunnel exit and with a trajectory away from the metal surface that would go deeper inside the metal. Therefore, all solutions with $\beta=1$ are excluded here. All pair contributions of the main manuscript have ($\beta$,\,$m$)=($-1$,0).

Both saddle points that form one saddle-point pair c[x]($\beta$,\,$m$) can be distinguished with the additional index $\alpha$ ($\alpha=-1$ for long orbits and $\alpha=1$ for short orbits) and addressed in the notation c[x]($\alpha$,\,$\beta$,\,$m$)~\cite{Milosevic2002_classification}. Here, we are interested in the saddle-point pair contributions c[x]($\beta$,\,$m$) that build up the photoemission spectrum.

In the c[x]($\beta$,\,m) classification scheme, the relevant saddle-point pair contributions of the main manuscript read $\mathrm{C1}=\mathrm{c1}(-1,0)$, $\mathrm{C2}=\mathrm{c2}(-1,0)$, and $\mathrm{C3}=\mathrm{c3}(-1,0)$. The red saddle-point pair contributions in Fig.~S\ref{fig:alleTraj}(c) are longer trajectories ($m>0$) born in the main cycles of the pulse (c1, c2, c3) at negative electric field ($\beta=-1$). For short pulses, long trajectories have either low spectral weight (ionization before the main cycles of the pulse at low electric field strength) or low cutoff energies (rescattering at low amplitude of the pulse envelope).

Consequently, all red pair contributions in Fig.~S\ref{fig:alleTraj}(c) are negligible in the plateau region ($E_{\mathrm{f}} > 2\,U_{\mathrm{p}}$). Only c1($-1$,1), born in cycle~1 and rescattering in cycle~3 is close to being relevant. It has a cutoff at $6.2\,U_{\mathrm{p}}$ in Fig.~S\ref{fig:alleTraj}(c) that with the near-field-induced energy shift of $1.0\,U_{\mathrm{p}}$ (4.0\,eV) would appear at $5.2\,U_{\mathrm{p}}$. In Fig.~S\ref{fig:alleTraj}(a), c1($-1$,1) is responsible for the small deviations of the red spectrum from the blue spectrum at energies above $2\,U_{\mathrm{p}}$. We conclude that c1($-1$,1) and all the other red pair contributions shown in Fig.~S\ref{fig:alleTraj}(c) can be disregarded due to their negligible spectral weight.

The trajectory of an electron is determined by the electric field of the laser pulse. If the trajectory of an electron born at negative electric field is longer than one optical cycle, it can enter the negative half space and be lost inside the metal. Whether this will happen depends on the ionization time $t_0$, the exact shape of the electric field after ionization, and, most importantly, the tunnel exit and therefore the electric field strength at $t_0$. A lower electric field strength at $t_0$ leads to a higher tunnel exit and makes it more likely that the electron trajectory will remain in the positive half space.

In Fig.~S\ref{fig:alleTraj}(d), electron trajectories are shown for the long (red) and the short (blue) orbit of the pair contribution c1($-1$,1) for a final energy of $E_{\mathrm{f}}=1.5\cdot10^{-5}\,U_{\mathrm{p}}$ (smallest energy in the simulation; solid lines) and $14.9\,U_{\mathrm{p}}$ (highest energy in the simulation; dotted red and dash-dotted blue lines) together with the electric field and the vector potential. The metal-vacuum interface is indicated with a solid black line. The electron trajectories are most different at low final energies and approach each other at energies above the cutoff energy, sampling the exact form of the saddle points (see Fig.~2(b) of the main manuscript for c1($-1$,0) for the energy trend).

Most importantly, for the parameters of the experiment and $\mathrm{CEP}=0$, all electron trajectories of c1($-1$,1) in the quantum orbit simulation enter the metal surface. These electrons are lost inside the tip and c1($-1$,1) will not contribute to the spectrum. This behavior is typical for long trajectories for the parameters of the experiment. The exclusion of these long electron trajectories for the calculation of the photoemission spectrum is therefore often not only convenient because of their low spectral weight, but also necessary because they enter the negative half space.

In the main manuscript, only the short pair contributions with ($\beta$,\,$m$)=($-1$,0) of the laser cycles c1 to c4 are included, and all pair contributions with longer trajectories ($m>0$) are excluded. More precisely, for the CEP map in Fig.~3(a) of the main manuscript, simulations at 16 evenly spaced CEP values in the range $[0; 1.875 \pi]\,\mathrm{rad}$ were used. These include C2 and C3 for all CEPs, C1 for CEPs in the range $[0; 0.75 \, \pi]\,\mathrm{rad}$ and $\mathrm{C4}=\mathrm{c4}(-1,0)$ for CEPs in the range $[0.625 \pi; 1.875 \pi]\,\mathrm{rad}$. Apart from these, no other contributions have been used in Fig.~3(a) of the main manuscript. C1 is completely negligible for $\mathrm{CEP}>0.75 \pi \, \mathrm{rad}$ and the same is true for C4 for $\mathrm{CEP}<0.625 \pi \, \mathrm{rad}$.

In Figs.~4(b),(c) of the main manuscript, the photoemission spectrum from the quantum orbit simulation is plotted together with the pair contributions of the short orbits C2, C3 and C4 for interpretation. For the same purpose, the pair contributions C1, C2 and C3 are plotted in the CEP maps of Figs.~3 and ~4(a) of the main manuscript. An easy interpretation of the full spectrum in this way is only possible if the pair contributions do not lead to unexpected and strong interference effects among each other, except from photon order peaks.

In Fig.~S\ref{fig:alleTraj}(b), the shortest pair contributions C1, C2, and C3 are shown in blue, together with their coherent sum in yellow. The yellow spectrum is the same as the blue spectrum in Fig.~S\ref{fig:alleTraj}(a), but it has been smoothed to remove photon order peaks. Fig.~S\ref{fig:alleTraj}(b) also shows the incoherent sum of the saddle-point pair contributions $p \left(|M_{\mathrm{C1S}}+M_{\mathrm{C1L}}|^2+|M_{\mathrm{C2S}}+M_{\mathrm{C2L}}|^2+\dots\right)$ (dash-dotted black line) and the incoherent sum of the isolated saddle-point contributions $p \left(\sum_{i=1}{|M_i|^2}\right)$ (red dotted line; like the black dotted line in Fig.~2(c) of the main manuscript). Notably, the dash-dotted black line closely matches the yellow line, even at $1.8 \, U_{\mathrm{p}}$, $3.1 \, U_{\mathrm{p}}$, $4.4 \, U_{\mathrm{p}}$, $6.0 \, U_{\mathrm{p}}$, and $9 U_{\mathrm{p}}$, where c2($-1$,0) and c1($-1$,0) have approximately equal spectral weights and strong photon order peaks appear (see Fig.~S\ref{fig:alleTraj}(a)). The same behavior is true for all CEPs for the parameters of the experiment, at least if only the short saddle-point solutions (($\beta$,\,$m$)=($-1$,0)) are taken into account.

In the inset of Fig.~S\ref{fig:alleTraj}(a), the red photoemission spectrum (including all pair contributions of Fig.~S\ref{fig:alleTraj}(c)) is smaller than the blue photoemission spectrum (including only the blue pair contributions in c), indicating destructive interference. This can be attributed to the presence of the saddle-point pair c1($-1$,1). Because c1($-1$,1) has a smaller spectral weight than c2($-1$,0) and c1($-1$,0), the effect is rather small. Due to the nonlinearity of the photoemission process, spectral weights of different pair contributions often differ by more than one magnitude. This makes interference effects among them less severe. Plotting the photoemission spectrum together with the isolated pair contributions for interpretation is always reasonable. A more careful investigation is necessary only if unexpected interference effects appear. In the most general case, interference effects in photoemission spectra can be investigated by switching on and off spectral contributions in the simulation.

\section{Difference map}
\label{app:Diffmap}

\begin{figure*}[h!]
\includegraphics[width=15cm]{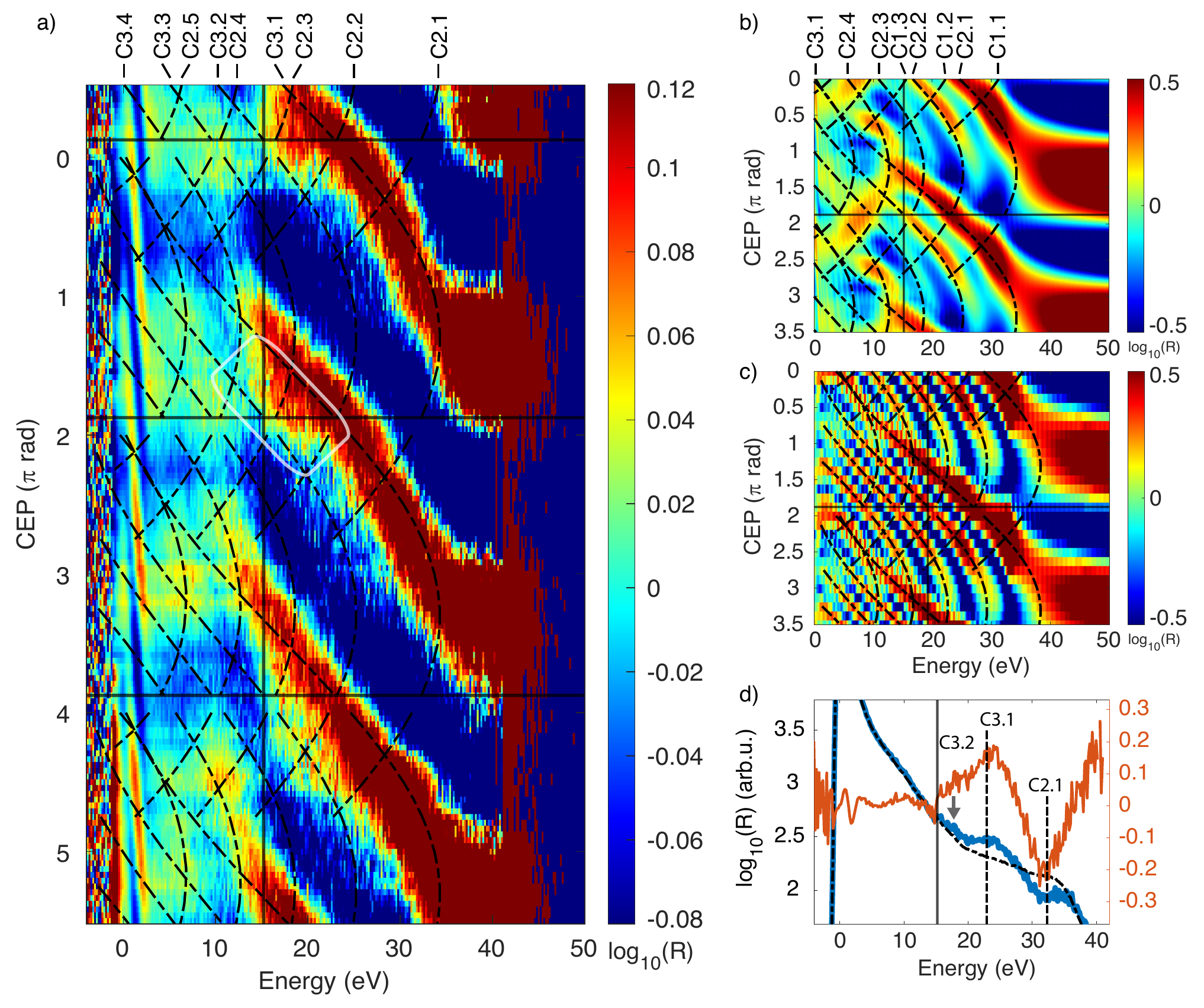}
\caption{\label{fig:Diffmap}Difference maps of our experimental photoemission spectra (a), TDSE simulation (b), and quantum orbit simulation (c) with the intracycle interference maxima of the quantum orbit simulations overlaid on them (dash-dotted lines; same as in Figs.~3 and 4(a) in the main text). (d) Difference spectrum of the experimental data at a CEP of $1.875 \pi \, \mathrm{rad}$ (horizontal black line in (a)). The energy position for the expected intracycle interference maximum C3.2 for this CEP is marked with a vertical solid line in (a)--(d). The difference spectrum results from the subtraction of the CEP-averaged spectrum (black) from the full spectrum (blue; see (d)). The intracycle interference maxima C2.1 and C3.1 can be spotted with high contrast for most CEPs. Inherently to the data representation, C2.1 is not visible around $\mathrm{CEP}=2\,\pi$ in the difference map, although being clearly present in the full data and simulations (see also (d)). Remarkably, there is a signal around the expected intracycle interference positions of C3.2 and C2.2 (C3.2 becomes C2.2 at $\mathrm{CEP} = 2 \pi \, \mathrm{rad}$). This signal is most pronounced in the CEP region between $1.5 \pi$ and $2.1 \pi \, \mathrm{rad}$ and energy region between 13 to 22\,eV (white box). It extends to C3.1 in the form of a small plateau, which is also apparent in (d). Traces of signal can be further tracked around C2.2 up to a CEP of $3.4 \pi \, \mathrm{rad}$. Besides of this, the experimental difference map (a) shows additional features below $13 \, \mathrm{eV}$, where the direct electrons dominate. In (b) and (c), smoothing in the energy domain has been applied. See text for more details.}
\end{figure*}

The appearance of the maxima C2.1 and C3.1 in our experimental data (see Figs.~1(b) and 4(a) in the main text) clearly originates from intracycle interference. According to our fully coherent TDSE and quantum orbit simulations, further intracycle interference maxima are predicted at lower energies. However, as the travel time difference between short and long orbits increases towards lower energies, electronic decoherence in the metal hinders the observation of these signals, as outlined in the main text and our phenomenological model for the electronic decoherence. To clarify whether further intracycle interference maxima are present in our experimental photoemission spectra, in addition to C2.1 and C3.1, we calculate difference maps. In the difference map, the CEP-averaged photoemission spectrum, i.e., the part of the spectrum that is not CEP modulated, is subtracted from the CEP-resolved photoemission spectra. Therefore, small CEP-dependent modulations are easier to trace in the difference map than in the CEP map of the full spectra that have a significant signal background (see Fig.~4(a) in the main text). In Fig.~S\ref{fig:Diffmap}(a), we show the difference map of our measurement data, together with the corresponding difference maps of our TDSE (b) and quantum orbit simulations (c) for comparison.

The experimental difference map shows various interesting features. The most prominent is the strong modulation in the plateau region in the form of red and blue stripes at energies above 12\,eV and is linked to the intracycle interference C2.1 and C3.1 (dash-dotted lines). The dash-dotted lines mark the positions of the quantum orbit expectations for the intracycle interference maxima (same as in Fig.~4(a) in the main text). It should be noted that C2.1 is not directly visible in the difference spectra for CEPs in the range between  $1.5 \pi$ and $2 \pi\,\mathrm{rad}$. This is directly related to the definition of the difference map and is apparent in Fig.~S\ref{fig:Diffmap}(d). Fig.~S\ref{fig:Diffmap}(d) shows the experimental difference spectrum at a CEP of $1.875 \pi \, \mathrm{rad}$ (horizontal black lines in (a)--(c)), together with the corresponding (full) experimental spectrum (blue) and the CEP-averaged spectrum (dash-dotted line). The energy positions of the expected intracycle interference maxima C2.1, C3.1, and C3.2 are marked with vertical lines (dashed black lines for C2.1 and C3.1, and full black line for C3.2). The energy position of C3.2 for this CEP ($\mathrm{CEP} = 1.875 \pi \, \mathrm{rad}$) is also highlighted in (a)--(c) as a vertical black line. For this CEP, C2.1 does not appear as a peak in the difference spectrum because the spectral yield of C2.1 is below the CEP-averaged spectrum.

Remarkably, in Fig.~S\ref{fig:Diffmap}(d), a small hump is visible in the measurement data (blue) at an energy of $\sim 17.5\, \mathrm{eV}$ ($4.4 \, U_\mathrm{p}$), as indicated by the gray arrow. This hump appears at an energy slightly above the quantum orbit expectation for C3.2 at an energy of $15.2 \, \mathrm{eV}$ ($3.8 \, U_\mathrm{p}$) and is very similar to the hump in the spectrum at $\mathrm{CEP} = 1.75 \pi \, \mathrm{rad}$ in Figs.~4(b),(c) in the main text. In Fig.~S\ref{fig:Diffmap}(d), this hump leads to a plateau in the difference spectrum (red) between the quantum orbit expectations for C3.2 and C3.1. This plateau is also visible in the difference map in (a). The plateau-like appearance of the hump in the difference spectrum could be linked to the steep signal background, i.e., the CEP-averaged spectrum. The feature is most pronounced in the CEP region between $1.5\,\pi$ and $2.1 \pi \, \mathrm{rad}$ and the energy region between 13\,eV ($3.2 \, U_\mathrm{p}$) and 22\,eV ($5.5 \, U_\mathrm{p}$; white box in (a)). Weaker signals can be found around the quantum orbit expectation for C2.2 in the CEP region between $2.1 \pi$ to $3.4 \pi \, \mathrm{rad}$, up to an energy of 26\,eV ($6.4\,U_\mathrm{p}$). For reference, the (10-$U_\mathrm{p}$-)cutoff in our measurement appears at $8.6 \, U_\mathrm{p}$ (34\,eV) and not at around $10\,U_\mathrm{p}$, due to the optical near field. It should be noted that C2.2 turns to C3.2 at $\mathrm{CEP} = 2 \pi \, \mathrm{rad}$ in our notation.

So far, we have restricted our discussion to a CEP region of approximately one period of CEP change. Regarding the weak appearance of the signal and the unexpected decrease in signal amplitude after $\mathrm{CEP} \sim 2.1 \pi \, \mathrm{rad}$, the question arises if the feature is reproducible. In our experiment, we measured CEP-resolved strong-field photoemission spectra by changing the CEP over 3 periods. In Fig.~S\ref{fig:Diffmap}(a), we show the difference map of our complete data set. Horizontal lines are plottet at CEPs of $-0.125 \pi$ and $3.875 \pi \, \mathrm{rad}$ for comparison with the difference spectrum at a CEP of $1.875 \pi \, \mathrm{rad}$. Compared to $\mathrm{CEP}=1.875 \pi \, \mathrm{rad}$, the possible intracycle interference feature C3.2/C2.2 looks very similar in the CEP region around $-0.125 \pi \, \mathrm{rad}$, but slightly different in the CEP region around $3.875 \pi \, \mathrm{rad}$. In the CEP region around $3.875 \pi \, \mathrm{rad}$, the signal seems to fit even better the expectation of C2.2. However, since it deviates from the signals around $-0.125 \pi$ and $1.875 \pi \, \mathrm{rad}$, it could be connected to an instability in the measurement.

We note that the signals around C3.2/C2.2 are found in all three CEP periods of the measurement. They are real features, appear as humps in the measurement data around the expected intracycle interference maxima C3.2/C2.2 (see Figs.~4(b),(c) in the main text and Fig.~S\ref{fig:Diffmap}) and are most pronounced in the CEP region between $1.5 \pi$ and $2.1 \pi \, \mathrm{rad}$.

Here, it is worth comparing the difference map of our measurement (see Fig.~S\ref{fig:Diffmap}(a)) with those of our TDSE and quantum orbit simulations in Figs.~S\ref{fig:Diffmap}(b),(c). Apart from the intracycle interference signals C2.2 and C2.3, the difference maps of both simulations do not reveal any other features in the relevant CEP and energy regions. Similarly to the difference map of the experimental data, the signal amplitude of C2.2 in the difference maps of the simulations decreases towards higher CEPs in the CEP region between $2 \pi$  and $4 \pi \,\mathrm{rad}$, especially for $\mathrm{CEP} > 2.6 \pi \, \mathrm{rad}$. However, in contrast to our simulations, there is no strong signal around C2.2 in the CEP region between $2.1 \pi$ and $2.6 \pi \mathrm{rad}$.

The signal decrease of C2.2 is linked to the CEP shifting the electric field minimum of cycle 2 (see Fig.~2(a) of the main text) to earlier times and therefore lower pulse envelope amplitude. This corresponds to a lower ionization rate and, therefore, a lower spectral yield for C2. The effect can already be seen for C2.1 in the CEP map of the experimental data in Fig.~3 in the main text and all dominant intracycle interference maxima C2.[X] in the CEP maps of the simulation results (see Figs.~3 in the main text) but is less pronounced than in the difference map. We note that in the difference map, a local maximum that appears below the CEP average has a negative spectral yield and appears blue in the CEP map.

To conclude this discussion, we think that our observed signals around C2.2 and C3.2 in Fig.~S\ref{fig:Diffmap} indeed very likely originate from intracycle interference, because of the signal decrease towards higher CEP in the CEP region between $2 \pi$ and $4 \pi \,\mathrm{rad}$ and the absence of other features in the difference maps of our TDSE and quantum orbit simulations. The fact that our signals around C2.2 and C3.2 appear as plateaus in Fig.~S\ref{fig:Diffmap} rather than peaks may be explained by the steep signal background in conjunction with a reduced signal amplitude due to electronic decoherence and critical signal-to-noise ratio.

In the difference map of our experimental data in Fig.~S\ref{fig:Diffmap}(a), additional features appear at energies below 13\,eV ($3.2 \, U_\mathrm{p}$). In particular, there are features around the quantum orbit expectation of C2.4 in the energy region between 9\,eV ($2.2\,U_\mathrm{p}$) and 13\,eV ($3.2\,U_\mathrm{p}$) and CEP regions $\left[0.4 \pi, 1.1\pi\right]$, $\left[2.4 \pi, 3.1 \pi\right]$ and $\left[4.4 \pi, 5.1\pi \right]\,\mathrm{rad}$. However, it is highly doubtful that these features are really related to the intracycle interference C2.4, since the influence of direct electrons is significant at energies below 13\,eV. Slightly above 0\,eV, the NILES effect is visible~\cite{Heimerl2025}.

It should be noted that various features in the TDSE result in Fig.~S\ref{fig:Diffmap}(b) at energies below 13\,eV are to some extent artificial. They are linked to strong ATI peaks that in the smoothing procedure merged together. The smoothing procedure is also the reason why C2.3 appears with low signal in the TDSE and is connected to the presence of direct electrons. This is not the case for the quantum orbit result in Fig.~S\ref{fig:Diffmap}, which does not include direct electrons.

\section{Phenomenological model for electronic decoherence}\label{app:Decoh}

In our measurement, intracycle interference shows up in the form of C2.1 and C3.1. Weak signs of intracycle interference are also present around C2.2 and C3.2, as discussed above. The reason for the strong suppression of intracycle interference in our measurement may be directly linked to electronic decoherence in our metal nanotip sample, e.g., due to electron-electron and electron-phonon scattering.

Intracycle interference results from the interference of short and long orbits. Electrons on short (S) and long (L) orbits are ionized $\left(t_0\right)$ and rescattering $\left(t_1\right)$ at different times. In the time differences $\Delta t_0 = |Re[t_{0\mathrm{S}}]-Re[t_{0\mathrm{L}}]|$ and $\Delta t_1 = |Re[t_{1\mathrm{L}}]-Re[t_{1\mathrm{S}}]|$, electronic decoherence obscures the phase relation between the corresponding short and long orbits. This decreases the coherence between the corresponding short and long orbits and leads to a reduced contrast of the related intracycle interference signals.

\begin{figure*}[h!]
\includegraphics[width=15cm]{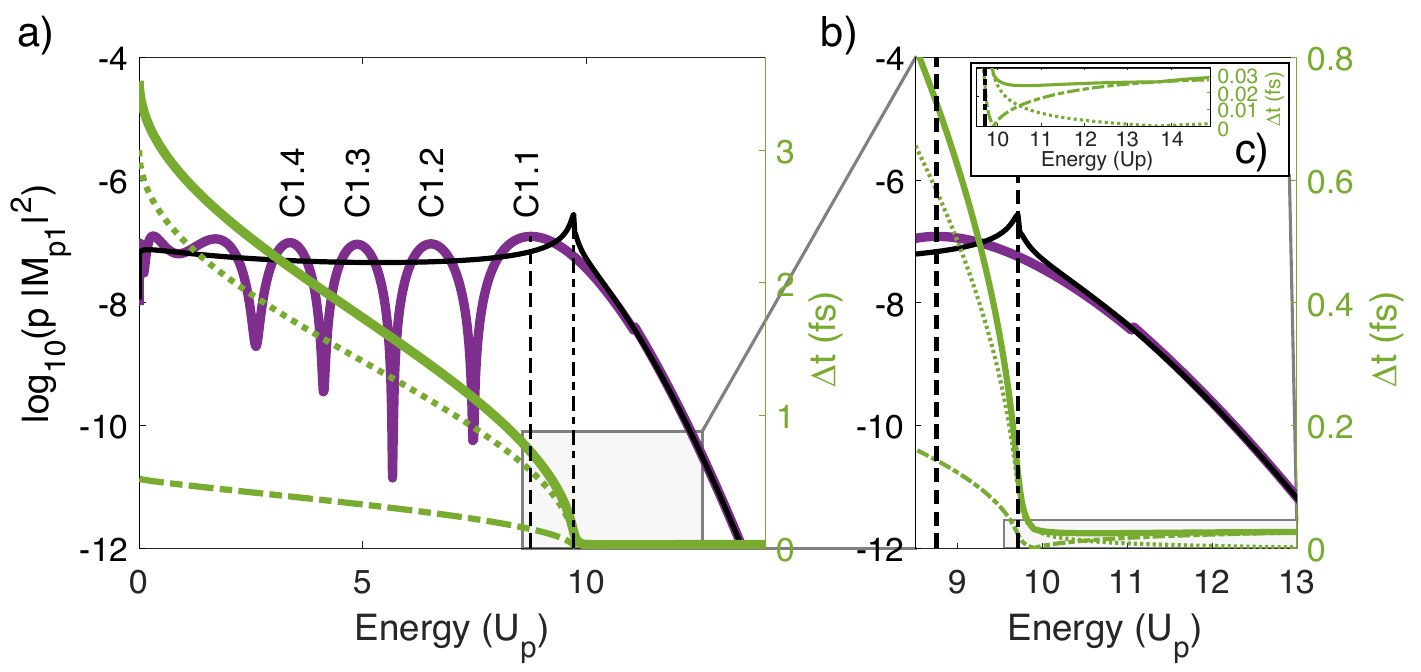}
\caption{\label{fig:TravelTimeDiff}(a): Intracycle interference contribution C1 at CEP~=~0 (purple) and the incoherent addition of the spectral weights of the corresponding short and long orbits $p \left(|M_\mathrm{S}|^2 +| M_\mathrm{L}|^2\right)$ (black; identical to Fig.~2(c) in the main text). The energy positions of C1.1 and of the cutoff, as evident by the artificial spike in the incoherent spectrum, are indicated by dashed and dash-dotted lines. The travel time difference between short and long orbits $\Delta t = \Delta t_0 + \Delta t_1$ (solid green line) is composed of the time difference in ionization times $\Delta t_0$ (dash-dotted green line) and of the time difference in rescattering times $\Delta t_1$ (dotted green line). $\Delta t$ amounts to $3.5 \, \mathrm{fs}$ (68 \% of the optical cycle) at energies close to zero and decreases towards the cutoff. It amounts to 1.4\,fs, 1.1\,fs, and 0.7\,fs at C1.2, at the rate minimum at $7.5 \, U_\mathrm{p}$ and C1.1, respectively. (b),(c): Insets of (a) in the cutoff region. (b): After C1.1, $\Delta t$ quickly drops to 0.12\,fs at the cutoff at $9.7 \, U_\mathrm{p}$ and reaches $\sim 26 \, \mathrm{as}$ at $10 \, U_\mathrm{p}$. (c): From $10 \, U_\mathrm{p}$ towards higher energies, $\Delta t$ stays roughly constant, which is due to the different energy dependencies of $\Delta t_0$ and $\Delta t_1$. See text for more details.}
\end{figure*}

At energies below the cutoff, electrons on long orbits are ionized before electrons on short orbits and are rescattering at later times, i.e., $Re[t_{0\mathrm{L}}] < Re[t_{0\mathrm{S}}]$ and $Re[t_{1\mathrm{S}}] < Re[t_{1\mathrm{L}}]$. In order to discuss the effects of electronic decoherence inside the metal on our photoemission spectra, we calculate $\Delta t_0$, $\Delta t_1$, and their sum $\Delta t = \Delta t_0 + \Delta t_1$. We refer to $\Delta t$ as the travel time difference between short and long orbits. $\Delta t_0$, $\Delta t_1$, and $\Delta t$ of the saddle-point pair contribution c1(-1,0), i.e., C1 of Fig.~2 in the main text, are exemplary shown in Fig.~S\ref{fig:TravelTimeDiff} as a function of the final energy, together with their coherent spectral contribution $p |M_\mathrm{S} + M_\mathrm{L}|^2$ and their incoherently added spectral weights $p \left(|M_\mathrm{S}|^2 +|M_\mathrm{L}|^2\right)$ (black line). The energy modulation of $p |M_\mathrm{S} + M_\mathrm{L}|^2$ stems directly from intracycle interference, as outlined in the main text.

As can be seen in Fig.~S\ref{fig:TravelTimeDiff}, the travel time difference $\Delta t$ is mainly composed of $\Delta t_1$ at energies below the cutoff. This is not surprising, given the appearance of the saddle-point pair in Figs.~2(a),(b) in the main text. Crucially, the travel time difference $\Delta t$ and its components $\Delta t_0$ and $\Delta t_1$ are energy dependent. At energies close to zero, $\Delta t$ amounts to 3.5\,fs and covers a good fraction (68 \%) of the 5.2\,fs long optical cycle. $\Delta t$, as well as $\Delta t_0$ and $\Delta t_1$, then monotonically decrease towards higher energies. $\Delta t$ reaches $1.4 \, \mathrm{fs}$ at C1.2 at $6.5 \, U_\mathrm{p}$, $1.1 \, \mathrm{fs}$ at $7.5 \, U_\mathrm{p}$ (rate minimum), $0.7 \, \mathrm{fs}$ at C1.1 at $8.75 \, U_\mathrm{p}$, and $0.12 \, \mathrm{fs}$ at $9.7 \, U_\mathrm{p}$ at the cutoff (artificial spike in the incoherent spectrum).

If electronic decoherence is a limiting process in the measurement, then an intracycle interference pattern may be best observed in the form of a peak at C[X].1, where $\Delta t$ is the smallest among all corresponding intracycle interference maxima. A peak at C[X].1 emerges with higher contrast if, in addition to C[X].1, the first interference minimum (destructive interference) at lower energies is present. Speaking of C1, C1.1 emerges with higher contrast, if the electrons on short and long orbits are still sufficiently coherent at $\Delta t = 1.1 \, \mathrm{fs}$ at $7.5 \, U_\mathrm{p}$. However, constructive interference at C1.1 alone is already enough for the formation of a peak, as can be seen in Fig.~S\ref{fig:TravelTimeDiff}(a). There, C1.1 appears with a higher spectral yield than the incoherent spectrum (black).

At energies above C1.1, the travel time difference quickly drops and reaches $0.12 \, \mathrm{fs}$ at $9.7 \, U_\mathrm{p}$ at the cutoff (see Fig.~S\ref{fig:TravelTimeDiff}(b). It should be noted that the relations $\mathrm{Re}[t_{0\mathrm{L}}]<\mathrm{Re}[t_{0\mathrm{S}}]$ and $\mathrm{Re}[t_{1\mathrm{S}}]<\mathrm{Re}[t_{1\mathrm{L}}]$ are strictly true only at energies below the cutoff. The relations reverse at $9.9 \, U_\mathrm{p}$ ($\Delta t_0=0$) and $13.7 U_\mathrm{p}$ ($\Delta t_1 = 0$; see Fig.~S\ref{fig:TravelTimeDiff}(c)). Above the cutoff energy, the overall travel time difference $\Delta t$ between short and long orbits stays roughly constant. For C1 in Fig.~S\ref{fig:TravelTimeDiff}(c), it amounts to $\sim 0.026 \, \mathrm{fs}$, with the smallest value being $0.024 \, \mathrm{fs}$ at $10.6 \, U_\mathrm{p}$.

To further discuss the effect of decoherence on the intracycle interference signals, we utilize a phenomenological model, similarly to~\cite{Heide2022}. First, we write the coherent sum in the spectral yield $p |\sum_{i=1}^{N}{M_i}|^2$ in the form of Eq.~\eqref{eq:decoh_1}, where $^{*}$ denotes the complex conjugate.

\begin{equation}
\label{eq:decoh_1}
\left|\sum_{i=1}^{N}{M_i}\right|^2 = \left(\sum_{i=1}^{N}{M_i}\right) \cdot \left(\sum_{j=1}^{N}M_j^{*}\right) = \sum_{i=1}^{N}{|M_i|^2}+\sum_{i=1}^{N}\sum_{j=1, j\neq i}^{N}{M_i M_j^{*}}
\end{equation}

In order to model decoherence, we then introduce a damping factor, which acts on the cross terms $M_i M_j^{*}$ ($i\neq j$). With this, Eq.~\eqref{eq:decoh_1} becomes:

\begin{equation}
\label{eq:decoh_2}
\left|\sum_{i=1}^{N}{M_i}\right|^2 \rightarrow \sum_{i=1}^{N}{|M_i|^2}+\sum_{i=1}^{N}\sum_{j=1, j\neq i}^{N}{M_i M_j^{*} \exp{\left( -\Delta t/T_2 \right)}}
\end{equation}

In our phenomenological damping factor, $T_2$ is the dephasing time, i.e., the electronic coherence time of our tip. Here, we do not distinguish between the composition of $\Delta t$, i.e., we assume the same $T_2$ for $\Delta t_0$ and $\Delta t_1$.

In Eq.~\eqref{eq:decoh_2}, we cannot use the uniform approximation. Therefore, we always have an artificial spike at the cutoff. In order to better discuss the effect of the decoherence term, we calculate the difference of our partially coherent spectrum to the incoherent spectrum, i.e., in terms of Eq.~\eqref{eq:decoh_2}:

\begin{equation}
\label{eq:decoh_2_Differenz}
p\left|\sum_{i=1}^{N}{M_i}\right|^2 - p \sum_{i=1}^{N}{|M_i|}^2 = p \sum_{i=1}^{N}\sum_{j=1, j\neq i}^{N}{M_i M_j^{*} \exp{\left( -\Delta t/T_2 \right)}}
\end{equation}

\begin{figure*}[h!]
\includegraphics[width=15cm]{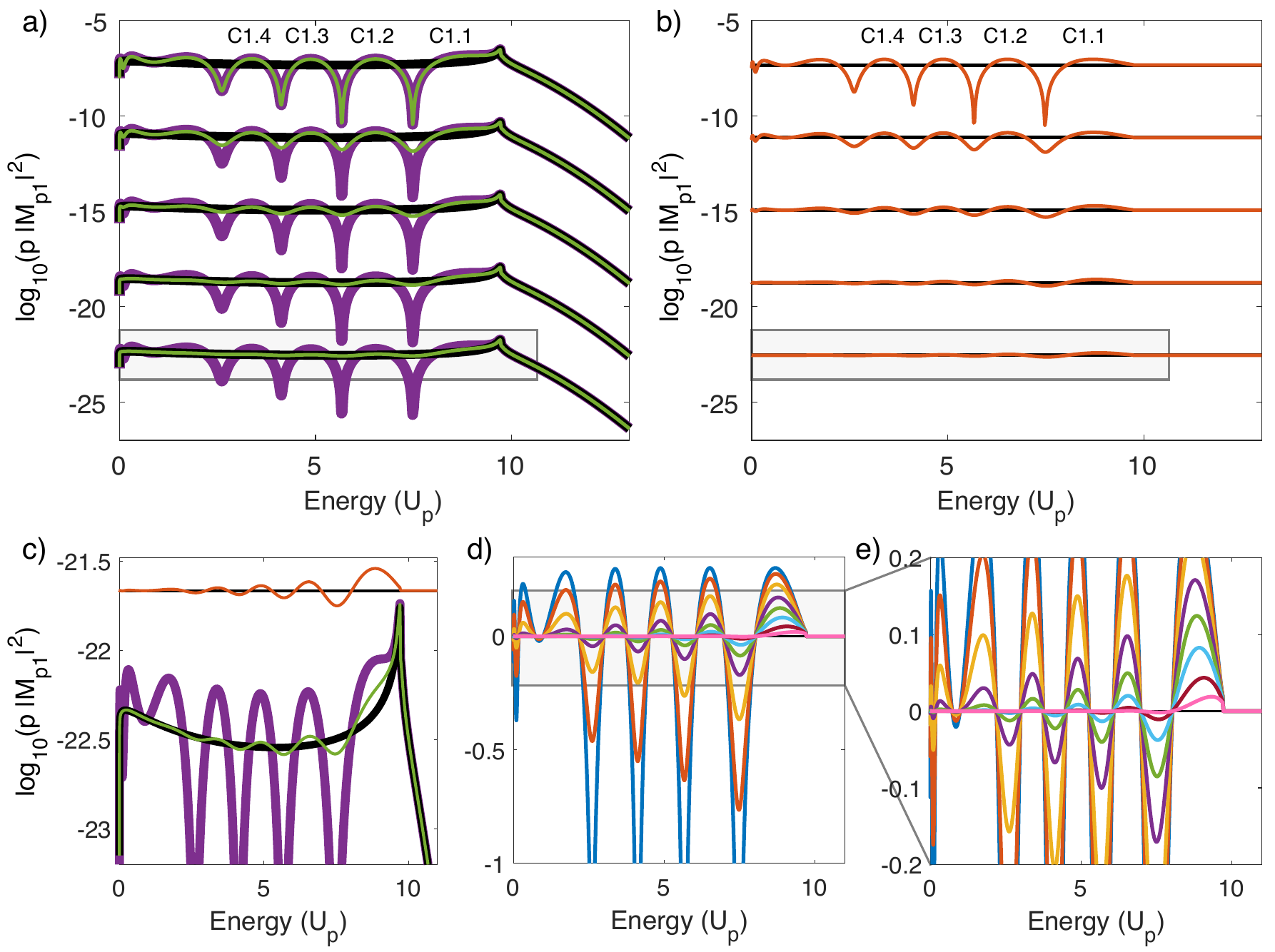}
\caption{\label{fig:Decoh1}(a): Simulated partially coherent spectral yield of C1 for various coherence times $T_2$ (green; $T_2$ in fs from top to bottom: $\infty$, 6, 2, 1, 0.65) together with the coherent and incoherent results (purple and black; same as in Fig.~S\ref{fig:TravelTimeDiff}). The spectra are shifted on the ordinate for better visualization. (b): Partially coherent spectral yield, as in (a), but with the incoherent spectrum (black) subtracted (red). Our model predicts a smooth transition between the fully coherent $\left(T_2 = \infty\right)$ and incoherent case $\left(T_2 = 0\right)$. (b) is particularly helpful for discussing the effects close to the cutoff, where an artificial spike occurs at $9.7 \, U_\mathrm{p}$ in (a). (c): Partially coherent spectrum for $T_2 = 0.65 \, \mathrm{fs}$ and corresponding spectrum with the incoherent spectrum subtracted (boxes in (a) and (b)), shifted on the ordinate for better comparison. For this coherence time, the damping factor leads to a strong suppression of the interference signal at energies below $7 \, U_\mathrm{p}$. This could explain why only C1.1 appears with significant amplitude in the measurement. (d): Partially coherent spectra with the incoherent spectra subtracted for various coherence times ($\infty$ (dark blue), 6\,fs (red), 2\,fs (yellow), 1\,fs (purple), 0.65\,fs (green), 0.45\,fs (light blue), 0.3\,fs (brown), 0.2\,fs (pink)). (e): Inset of (d). For very short coherence times $\left(T_2 \ll \Delta t_\mathrm{C1.1} = 0.5 \, \mathrm{fs}\right)$, also the intracycle interference maximum C1.1 gets strongly suppressed. See text for more details.}
\end{figure*}

Figures~S\ref{fig:Decoh1}(a) and~S\ref{fig:Decoh1}(b) show the partially coherent spectra of C1 for various damping constants $T_2$, and the corresponding spectra with the incoherent contribution subtracted. To the latter, we also refer to as the coherent part of the partially coherent spectrum. We note that the results for C1 are representative for all quantum orbit contributions c1(-1,0), i.e., C1, C2, and C3 for our two-cycle pulse. Small deviations arise due to slight differences in the form of the saddle points in the complex plane (see Figs.~2(a),(b) in the main text).

For $T_2 = \infty$, our model is identical to the fully coherent spectrum $\left(\exp{\left[ -\Delta t/T_2 \right]} =1 \right)$, while for $T_2 = 0$, it is identical to the fully incoherent spectrum $\left(\exp{\left[ -\Delta t/T_2 \right]} =0 \right)$. Our model predicts a smooth transition between both regimes as a function of the coherence time $T_2$.

In Fig.~S\ref{fig:Decoh1}(c), we show the partially coherent spectrum for $T_2 = 0.65 \, \mathrm{fs}$, with and without the incoherent spectrum subtracted. The latter is preferable for discussing decoherence effects because the signal background, particularly the artificial spike at the cutoff at $9.7 \, U_\mathrm{p}$, is subtracted. As the travel time difference $\Delta t$ of the interfering electrons already amounts to $1.4 \, \mathrm{fs} \gg T_2 = 0.65 \, \mathrm{fs}$ for C1.2, C1.2, and all subsequent maxima (see Fig.~S\ref{fig:TravelTimeDiff}) are strongly suppressed in Fig.~S\ref{fig:Decoh1}(c). For even shorter coherence times (see Figs.~S\ref{fig:Decoh1}(d),(e)), also the amplitude of C1.1 is significantly decreased $\left(\Delta t_\mathrm{C1.1} = 0.7 \, \mathrm{fs}\right)$.

\begin{figure*}[h!]
\includegraphics[width=13cm]{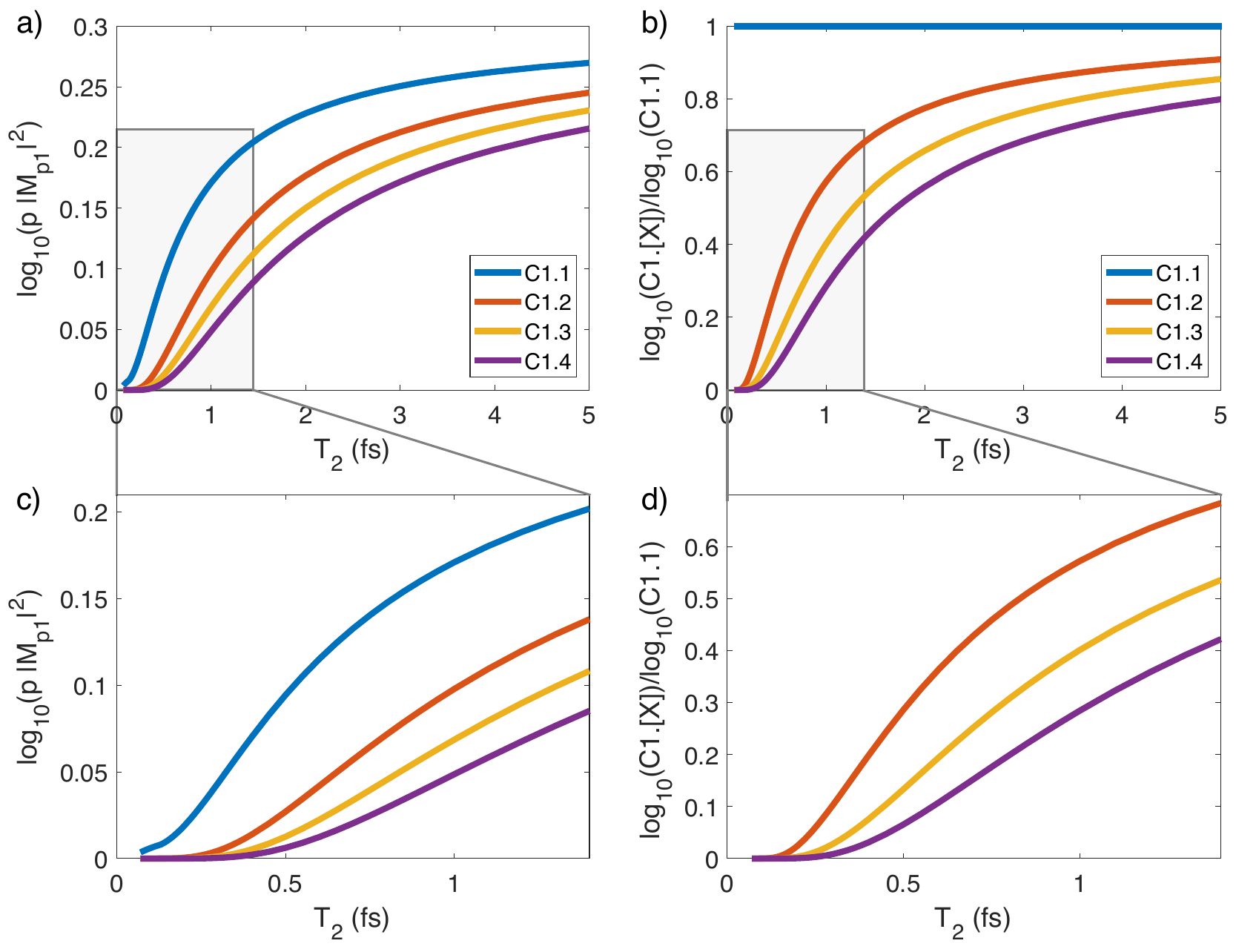}
\caption{\label{fig:T2_vs_rate}(a): Simulated spectral yields of the intracycle interference maxima C1.1, C1.2, C1.3, and C1.4 as a function of the coherence time $T_2$. (b): Ratios of the logarithmic rates of C1.[Y] to C1.1. (c),(d): Insets of (a) and (b). The rates of the intracycle interference maxima strongly decrease for $T_2<2\,\mathrm{fs}$. For the absolute amplitude strength of C[X].1 and the strong suppression of C[X].2 in our experimental data (see Figs.~4(c),(d) in the main text), our phenomenological model suggests coherence times in the range of $1\,\mathrm{fs}$ and below. See text for more details.}
\end{figure*}

Figs.~S\ref{fig:T2_vs_rate}(a),(c) show the spectral yields of the first four intracycle interference maxima, with the incoherent contribution subtracted (coherent parts of the partially coherent spectra), as a function of $T_2$.

From Figs.~S\ref{fig:T2_vs_rate}(a),(c), we see that the spectral yields of the intracycle interference maxima C1.[Y] strongly decrease for $T_2 < 2 \, \mathrm{fs}$. At a fixed $T_2$, the effect is stronger the higher the maximum number Y, i.e., C1.4 is most affected and C1.1 is less affected. At $T_2 = 1 \, \mathrm{fs}$, the amplitudes of C1.1, C1.2, C1.3, and C1.4 are 0.171, 0.098, 0.069, and 0.049 on logarithmic scale and the units of Fig.~S\ref{fig:Decoh1}, respectively. At $T_2 = \infty$, the corresponding amplitudes of C1.1--C1.4 are all $0.30$ on logarithmic scale (with small deviations at higher precision; see Fig.~S\ref{fig:Decoh1}(d)). These amplitudes are directly to be compared with the corresponding amplitudes in the measurement data. We note that the absolute amplitude decreases on logarithmic scale at $T_2=1 \,\mathrm{fs}$ correspond to relative decreases of 43, 67, 77, and 84\,\% for C1, C2, C3, and C4, respectively.

In our experimental data, we find very clear signatures of the intracycle interference maxima C[X].1, in particular C2.1 and C3.1, while C[X].2, i.e., C2.2 and C3.2 appear only as weak humps and traces on the difference map in of Fig.~S\ref{fig:Diffmap}(a). Therefore, it is illustrative to plot the relative rates C1.[Y] with respect to C1.1. These ratios are shown in Figs.~S\ref{fig:T2_vs_rate}(b),(d). At $T_2 = 1 \, \mathrm{fs}$, the logarithmic rate ratios C1.2/C1.1, C1.3/C1.1, and C1.4/C1.1 amount to 43, 60, and 72\,\%, respectively. At $T_2 = 0.65 \, \mathrm{fs}$, these ratios are 60, 78, and 87\,\%, respectively. The absolute signal strengths of C2.2 and C3.2 and the strong suppression of C2.2 and C3.2 in our measurement data (see Figs.~4(b),(c) in the main text) suggest coherence times in our tungsten tip in the range of $1\,\mathrm{fs}$ or below during the measurement, according to our phenomenological model.

\begin{figure*}[h!]
\includegraphics[width=12cm]{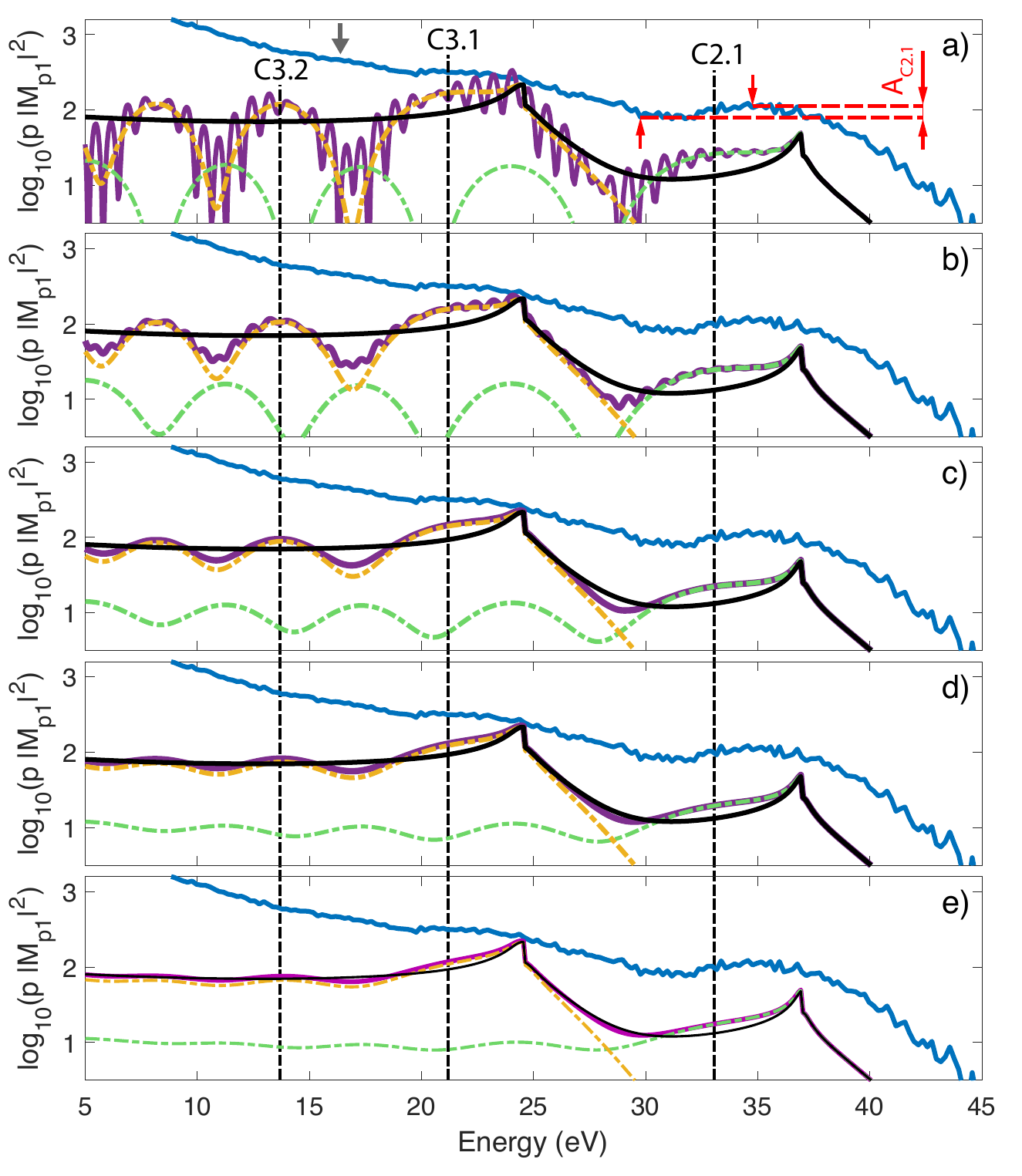}
\caption{\label{fig:Decoh2_mitExp}Simulated partially coherent spectral yields for our pulse parameters and a CEP of $1.75 \, \pi \, \mathrm{rad}$ for $T_2 = \infty$ (a), 6\,fs (b), 2\,fs (c), 1\,fs (d), and 0.65\,fs (e), according to our phenomenological model (purple). The individual contributions of C2 (green) and C3 (yellow) are shown as dash-dotted lines, and the experimental spectrum is shown in blue. Color and linewidth have been slightly adjusted in (d). As the coherence time $T_2$ decreases from (a) to (d), both intercycle and intracycle interference become more strongly suppressed. While ATI peaks are already strongly suppressed in (b) for $T_2 = 6 \, \mathrm{fs}$ and no more visible in (c) ($T_2=2 \, \mathrm{fs} < T_\mathrm{opt}$), the intracycle interference signal is still visible in (d) for $T_2=0.65 \, \mathrm{fs}$. The deviations from the incoherent spectrum are particularly strong at C2.1 and C3.1. Strong suppression of C3.2 in our measurement suggests coherence times on the order of 1\,fs or below. See text for more details.}
\end{figure*}

We note that our discussion for C1 is also representative for C2 and C3 due to similar appearance of the saddle points in the complex time (see Figs.~2(a),(b) in the main text). The photoemission spectra for our two-cycle pulses are determined by C1, C2, and C3. The influence of C1, C2, and C3 strongly depends on the CEP, as outlined in the main text. For CEPs in the range between $0.5 \pi$ and $1.2 \pi \, \mathrm{rad}$, only C2 is dominant in the plateau region of the spectrum, while for other CEPs, two intracycle contributions are relevant. The latter gives rise to intercycle interference, i.e., the appearance of ATI peaks in the spectrum if the contributions are sufficiently coherent. The photoemission spectra are especially interesting for CEPs close to $2 \pi \, \mathrm{rad}$. There, C2.1 and C3.1 are directly visible in the plateau region, as well as traces of C2.2 and C2.3 (see Fig.~4 in the main text and Fig.~S\ref{fig:Diffmap}). Because of similar spectral yields of C2.1 and C3.1, our fully coherent quantum orbit and TDSE simulations also predict ATI peaks in this CEP region. In Fig.~S\ref{fig:Decoh2_mitExp}, we show the results of our phenomenological model for various coherence times $T_2$ at a CEP of $1.75 \, \pi \, \mathrm{rad}$, together with the experimental spectrum, similarly to Fig.~4(c) in the main text.

Fig.~S\ref{fig:Decoh2_mitExp} shows a strong decrease of the intracycle interference signals with increasing electronic coherence time $T_2$. We note that the prominence of C2.1, i.e., the amplitude $A_\mathrm{C2.1}$ roughly matches the corresponding deviations of the simulated partially coherent spectrum from the incoherent spectrum in (d) for $T_2 = 0.65 \, \mathrm{fs}$. There, also C3.2 is significantly suppressed compared to C3.1, which matches our experimental observation, i.e., very weak signals at the energy position of C3.2. C3.2 may be the origin of the small hump at an energy of $\sim 16 \, \mathrm{eV}$ in Fig.~S\ref{fig:Decoh2_mitExp}(e), as indicated by a gray arrow (see also Figs.~4(b),(c) in the main text). We note that this hump is reproducible but significant only for CEPs close to $2 \pi \, \mathrm{rad}$ (see Fig.~S\ref{fig:Diffmap}(a)), as discussed above. We note that only for coherence times in the range of 1\,fs or below, our phenomenological damping model can reproduce the absolute signal strengths of C[X].1 (e.g., C3.1 in Fig.~S\ref{fig:Decoh2_mitExp}) and the strong suppression of C[X].2 (C3.2 in Fig.~S\ref{fig:Decoh2_mitExp}) with respect to C[X].1 in our experimental data.

There are no ATI peaks visible in our experimental data (see Figs.~S\ref{fig:Decoh2_mitExp},~S\ref{fig:Diffmap}(a), and Figs.~2(b) and 4 in the main text). For coherence times of much less than one optical cycle ($T_\mathrm{1550} = 5.2 \, \mathrm{fs}$), this is expected, since ATI peaks originate from similar electron contributions from different cycles. The absence of ATI peaks in our measurement is also perfectly supported by our phenomenological model (see Fig.~S\ref{fig:Decoh2_mitExp}). In Fig.~S\ref{fig:Decoh2_mitExp}, the ATI peaks are already strongly suppressed for $T_2 = 6 \, \mathrm{fs}$ (not much above the noise level) and completely absent for $T_2 = 2 \, \mathrm{fs}$. However, we note that the travel time difference as it enters our phenomenological model is an artificial time difference for the ATI peaks, since the time the electron spends in the vacuum does not enter. Consequently, care has to be taken regarding ATI peaks in our phenomenological model.

Actually, ATI peaks have already been experimentally observed in strong-field photoemission from metal tips, but only for (slightly) longer laser pulses in terms of optical cycles~\cite{Schenk2010, Krueger2011, Bionta2013, Krueger2018, Heimerl2023} and mostly at $\lambda = 800 \, \mathrm{nm}$. In these measurements, the electronic coherence time must have been at least in the range of the optical cycle $T_{800} = 2.7 \, \mathrm{fs}$.

In addition to the coherence time, a minimum laser pulse duration is crucial for the emergence of ATI peaks. Broadly speaking, we only find ATI peaks if there are at least two contributions one optical cycle apart with similar spectral weights. In our case here, the spectral weights of C1, C2, and C3 are too different (see, e.g., Figs.~S\ref{fig:Decoh2_mitExp} and 4~(b),(c) in the main text for one CEP). Remarkably, in~\cite{Kuebel2021}, weak onsets of ATI peaks can be found in a strong-field experiment on atoms in the gas phase ($T_2$ should be larger than $T_\mathrm{opt}$) for only slightly longer pulses in terms of optical cycles. We note that we have also measured weak onsets of ATI peaks in a strong-field photoemission measurement of a gold nanotip at $\lambda = 1550 \, \mathrm{nm}$ for a seemingly similar laser pulse duration, i.e., 10\,fs, but at much lower laser intensity. Heating of the nanotip at higher laser intensities and the tip material may have a significant influence on the coherence time. For a given material, longer pulses are better for the emergence of ATI peaks, as there are more contributions that lead to the formation of ATI peaks.

In addition, we note that it may be hard to extract quantitative estimations for the coherence time from ATI peaks because their appearance is affected by the laser pulse duration in a complex way, i.e., regarding the number of contributing electron wave packets and also their spectral weights. In terms of absolute signal amplitudes, sharp ATI peaks may be easier to detect than the broad energy modulation of an intracycle interference pattern, which is more likely to be lost in the noise of the measurement. Care should be taken when comparing the absolute signal amplitudes of ATI peaks with those of intracycle interference signals. We will need to leave it to future work to investigate the appearance of ATI peaks and intracycle interference as a function of pulse duration and, intriguingly, coherence time. For short coherence times in the range of a few femtoseconds or below, (sub-) two-cycle driving pulses and {\it intracycle} interference will be the method of choice to yield quantitative insights to electron coherence.

The emergence of ATI peaks in the strong-field photoemission measurements in~\cite{Schenk2010, Krueger2011, Bionta2013, Krueger2018, Heimerl2023} suggests that the coherence time in these measurements was long enough to allow intracycle interference. However, although intracycle interference may well have occurred in these measurements, its detection was impossible. For the clear emergence of an intracycle interference signal in a strong-field photoemission measurement, two criteria need to be fulfilled. Firstly, the laser pulse duration needs to be on the order of two optical cycles or shorter. Secondly, the 10-$U_p$ cutoff needs to appear at a sufficiently high energy to resolve the energy modulation of the intracycle interference pattern.

The energy modulation of the intracycle interference pattern is directly related to the difference in rescattering times $\Delta t_1$ via $\Delta E = 2\pi/\Delta t_1$~\cite{Milosevic2006} and is therefore fixed in absolute energy (eV) by the laser wavelength. To a good approximation, all intracycle interference contributions differ only in terms of spectral yield (given by the electric field strength at $t_0$) and the cutoff energy (given by the envelope of the pulse at $t_1$).

If the pulse duration is significantly longer than two optical cycles, spectral contributions from different cycles appear with similar spectral yields but different cutoff energies. The intracycle interference patterns of these contributions in sum lead to a lower modulation depth. As a consequence, the combined intracycle interference pattern may not be detected in the measurement. The intracycle interference patterns are completely smeared out if there are multiple contributions with similar spectral yield that are shifted by a fraction of the energy modulation of the intracycle interference pattern.

We note that many strong-field photoemission experiments have been performed with pulses at $\lambda= 800 \, \mathrm{nm}$ that should have been short enough for the emergence of intracycle interference patterns (i.e., with pulse durations in the range of $2 T_\mathrm{opt} = 5.4 \, \mathrm{fs}$). However, in most strong-field experiments, the second criterion, i.e., a sufficiently high 10-$U_\mathrm{p}$ cutoff energy, is typically not met. For $\lambda = 1550 \, \mathrm{nm}$, the energy periodicity of the intracycle interference pattern (i.e., the energy difference between adjacent maxima) is approximately 8 to 9\,eV between C[X].1 and C[X].2, and 6\,eV between C[X].2 and C[X].3. For $\lambda = 800 \, \mathrm{nm}$, it is roughly twice that. In~\cite{Schenk2010, Krueger2011, Bionta2013, Krueger2018}, the 10-$U_\mathrm{p}$ cutoffs appear at too low energy to cover the energy modulation of the intracycle interference. We emphasize that, for very low cutoff energies in the range of one period of the intracycle interference pattern (as in \cite{Schenk2010, Krueger2011, Bionta2013}), the first minimum between C[X].1 and C[X].2 differs from the above estimate, yet it still appears in the low-energy part of the direct electrons and cannot be measured. Notably, in~\cite{Heimerl2023}, the 10-$U_\mathrm{p}$ cutoff appears at sufficiently high energy for the measurement at $\lambda = 800\,\mathrm{nm}$, but the pulse duration ($12 \, \mathrm{fs} \gg 2 T_\mathrm{opt} = 5.4 \, \mathrm{fs}$) was too long for the clear emergence of an intracycle interference pattern.

In our experimental data at $\lambda = 1550 \, \mathrm{nm}$, the 10-$U_\mathrm{p}$ cutoff appears at $\sim 35 \, \mathrm{eV}$ (see Figs.~1(b) and 4(a) in the main text). For the emergence of a similar intracycle interference pattern (number of intracycle interference periods) at $\lambda = 800 \, \mathrm{nm}$, the 10-$U_\mathrm{p}$ cutoff needs to be twice as high, i.e., it needs to appear at $\sim 70 \, \mathrm{eV}$. For this, the near-field intensity needs to be roughly 8 times higher than in our case, i.e., it needs to be $I_{\mathrm{NF}} = 1.4 \cdot 10^{14} \, \mathrm{W/{cm}^2}$). A factor of 4 arises from the frequency dependence of $U_\mathrm{p}$ ($U_\mathrm{p} = e^2 E_0^2/\left(4 m \omega_\mathrm{opt}^2\right) \propto I/\omega_\mathrm{opt}^2$).

We want to conclude this section with some remarks about electronic decoherence in HHG and our phenomenological damping model. In~\cite{Heide2022}, electron-hole coherence was experimentally investigated in a semiconductor material via the effect of photodoping with incoherent electron-hole pairs on HHG signals. The procedure described in~\cite{Heide2022} may be applicable to different semiconductor materials with sufficient HHG yield and can be used to determine their coherence time $T_2$. In contrast, strong-field photoemission from nanometric needle tips is, in principle, applicable to all solid materials that can be formed into a nanometric needle tip or mounted on the apex of a sample tip. While the travel time difference $\Delta t$ seems to be the relevant timescale for electronic decoherence in the investigation of intracycle interference, in semiconductor HHG it is the time the electrons and holes spend in the conduction and valence bands between ionization and recombination. The latter might be more limiting in terms of accessible time differences.

We note that, in terms of electronic decoherence, the rescattering process might be more complex than covered from our phenomenological model. Our phenomenological model could be adjusted accordingly, e.g., by using two dephasing times $T_2$ (damping factor in the form $\exp{\left(-\Delta t_0/T_{2,0}-\Delta t_1/T_{2,1}\right)}$) or by various other modifications.

A more realistic treatment of electronic decoherence in our quantum orbit simulations, e.g., resulting from electron-electron and electron-phonon scattering, would require the use of an open quantum system description~\cite{Caldeira1983, Ingold2002, Weiss2008, Breuer2002}. In particular, our system could be modeled within the framework of a Caldeira-Leggett model. In this approach, the system is coupled to an environment represented by a bath of harmonic oscillators.

\putbib[references.bib]
\end{bibunit}

\end{document}